\documentclass[a4paper,11pt]{article}
\pdfoutput=1 

\usepackage{jheppub} 

\usepackage[T1]{fontenc} 

\newcommand{\lr}[1]{\left(#1\right)}
\newcommand{\lrsq}[1]{\left[#1\right]}

\newcommand{\xx}{\ensuremath{\mathfrak{X}}}
\newcommand{\xxsp}{\ensuremath{\mathfrak{X}^{1/2}}}
\newcommand{\xxsm}{\ensuremath{\mathfrak{X}^{-1/2}}}
\newcommand{\nablah}{\ensuremath{\nabla^h}}
\newcommand{\ja}{{\ensuremath{J_5}}}

\title{\boldmath Spontaneous Lorentz symmetry breaking and one-loop effective action in the metric-affine bumblebee gravity}

\author[a]{Adri\`a Delhom,}
\author[b]{T. Mariz,}
\author[c]{J. R. Nascimento,}
\author[d,e]{Gonzalo J. Olmo,} 
\author[c]{A. Yu. Petrov,}
\author[c,1]{Paulo J. Porf\'{\i}rio, \note{Corresponding author}} 

\affiliation[a]{Laboratory of Theoretical Physics, Institute of Physics, University of Tartu, W. Ostwaldi 1,\\
50411 Tartu, Estonia}
\affiliation[b] {Instituto de F\'{i}sica, Universidade Federal de Alagoas,\\
57072-270, Macei\'{o}, Alagoas, Brazil}
\affiliation[c]{Departamento de F\'{\i}sica, Universidade Federal da 
Para\'{\i}ba,\\
 Caixa Postal 5008, 58051-970, Jo\~ao Pessoa, Para\'{\i}ba, Brazil}
 \affiliation[d]{Departament de F\'{i}sica Te\`{o}rica and IFIC, Centro Mixto Universitat de
	Val\`{e}ncia - CSIC,\\
	Universitat de Val\`{e}ncia, Burjassot-46100, Val\`{e}ncia, Spain}
\affiliation[e]{Universidade Federal do Cear\'a (UFC), Departamento de F\'isica,\\ Campus do Pici, Fortaleza - CE, C.P. 6030, 60455-760 - Brazil}

\emailAdd{adria.delhom@gmail.com}
\emailAdd{tmariz@fis.ufal.br}
\emailAdd{jroberto@fisica.ufpb.br}
\emailAdd{gonzalo.olmo@uv.es}
\emailAdd{petrov@fisica.ufpb.br}
\emailAdd{pporfirio@fisica.ufpb.br}

\abstract{The metric-affine bumblebee model in the presence of fermionic matter minimally coupled to the connection is studied. We show that the model admits an Einstein frame representation in which the matter sector is described by a non-minimal Dirac action without any analogy in the literature. Such non-minimal terms involve unconventional couplings between the bumblebee and the fermion field. We then rewrite the quadratic fermion action in the Einstein frame in the basis of 16 Dirac matrices in order to identify the coefficients for Lorentz/CPT violation in all orders of the non-minimal coupling $\xi$. The exact result for the fermionic determinant in the Einstein frame, including all orders in $\xi$, is also provided. We demonstrate that the axial contributions are at least of second order in the perturbative expansion of $\xi$. Furthermore, we compute the one-loop effective potential within the weak field approximation. }

\begin{document} 
\maketitle
\flushbottom

\section{Introduction}
\label{sec:intro}

It is well known that symmetries play an important role in physics. In particular, Lorentz symmetry  is key in describing all known physical theories and has been verified with a very high degree of precision (see, e.g.,~\cite{datatables}). Even so, in fundamental theories (string/M-theory), there are theoretical reasons that support that small violations of this symmetry may occur at the Planck scale~\cite{Kostelecky:1988zi, Kostelecky:1989jp, Kostelecky:1989jw, Kostelecky:1991ak, Kostelecky:1994rn}, with potentially observable effects even in the low-energy regime. From the experimental point of view, directly probing Planck-scale effects nowadays or, at least, for the foreseeable future is more than challenging. For this reason, the most reasonable approach  consists on studying effective theories where the effects of Lorentz symmetry breaking (LSB) at attainable energy scales are Planck-suppressed (see, for example, the discussion in~\cite{Georgi}).

In what concerns effective theories, our framework of reference is the so-called Standard-Model Extension (SME) \cite{Colladay:1996iz, Colladay:1998fq}. This phenomenological model describes the physics beyond the Standard Model (SM) based on the enlargement of the matter sector incorporating all possible Lorentz-CPT violating (LV) coefficients (which are supposed to be constants) coupled to the matter. Just in the same way, in \cite{Kostelecky:2003fs}, the gravitational sector was also incorporated into the SME in a Riemann-Cartan background. However, the LV/CPT coefficients, in curved space-time, instead of being constants, must be dynamical according to the long discussion in \cite{Kostelecky:2003fs, Jacobson:2000xp,KosLi}. The key point of this discussion is that constant vectors or tensors, in general, cannot be consistently defined in a generic curved space-time due to the so-called no-go constraints { (see, e.g.,~\cite{KosLi})}. As a consequence of all this, the consistent implementation of LSB in curved space remains an open question, for which the most compelling answer is that it must be introduced via {a spontaneous symmetry-breaking mechanism}. In this scenario, the LV/CPT coefficients arise as vacuum expectation values (VEV) from dynamical fields driven by non-trivial potentials. 

Among the models that consider LSB in curved space via the gravitational sector,  special attention have received the Einstein-Aether theory \cite{Jacobson:2000xp}, the Chern-Simons modified gravity \cite{Jackiw:2003pm}, and the bumblebee gravity model \cite{Kostelecky:2003fs}. The latter two theories, in particular, have been explored recently in a variety of different contexts, namely, in cosmology \cite{Mirzagholi:2020irt, Maluf:2021lwh}, black holes \cite{KumarJha:2020ivj, Casana:2017jkc, Gullu:2020qzu, Maluf:2020kgf}, and gravitational waves scenarios \cite{Bartolo:2017szm, Conroy:2019ibo,Boudet:2022wmb}. Moreover, degenerate higher-order scalar-tensor (DHOST) theories have been also deemed in LSB context \cite{Gao:2020qxy}. In many of the aforementioned models, the background geometry is fixed prior -- being (pseudo)-Riemannian. As argued in \cite{Kostelecky:2003fs}, a promising way forward consists in considering more generic geometric approaches where, apart from the metric, other degrees of freedom are taken into consideration. For example,  in \cite{Kostelecky:2003fs,ourtorsion}, the Riemann-Cartan geometry, within which the metric and torsion are treated as dynamical geometrical quantities, was considered  as the background in the gravitational sector of the SME, and in the context of the teleparallel gravity \cite{Li:2020xjt}. A natural generalization is the metric-affine (Palatini) approach in which the connection and metric are assumed independent geometrical entities, see { f.e.} \cite{Olmo:2011uz} and references therein. Along this line, \cite{Delhom:2019gxg, pj20} have been the first works to investigate the implications of the metric-affine approach in the LSB scenario.  In these papers, a metric-affine version of the bumblebee model was proposed and its classical and quantum aspects were explored. At the classical level, the field equations were derived and solved, as well as the weak field and post-Minkowskian limit in the presence of scalar and { spin-$1/2$} fields in the first-order perturbation of the non-minimal parameter $\xi$  were studied emphasizing the modifications of dispersion relations and the stability conditions. At the quantum level, these papers focused on one-loop corrections within the effective field theory approach { involving a} non-trivial VEV of the bumblebee field. It was shown that the propagators get corrections proportional to the VEV, thereby dramatically changing the one-loop level contributions in the corrections of the first order in $\xi$.       

In this paper, we investigate the effects of the metric-affine bumblebee model coupled to fermions without disregarding the gravitational effects. First, we consider generic $\xi$, e.g., obtaining non-perturbative result in $\xi$, and,  afterward, discuss its role within the LV context. Second, we will focus on one-loop divergent corrections via the Schwinger-DeWitt method, meanwhile restricting ourselves to first-order in $\xi$ due to technical reasons.  
We will also show that our results recover the ones found in \cite{pj20} at lowest order in $\xi$ by turning off the gravitational effects.  

The paper is organized as follows. We begin section \ref{sec:1} by giving a brief review of the metric-affine bumblebee model calling attention to the field equations and also their solutions. In section \ref{section3}, we discuss the vierbein formalism that is the suitable approach to investigate fermions in curved spaces. Moreover, we display the general form of the spin action in the Einstein frame and also we show the coefficients for Lorentz violation in the post-Minkowskian limit. Section \ref{section4} is devoted to  calculating the quantum corrections via functional methods. We then first set up the quadratic fermionic operator (a bilinear form of the fermion action) in the Einstein frame, and then we calculate the divergent part of the fermionic effective action by integrating it over the quantum fermion fields (noteworthy we will restrict our analysis to the situation where the fermion background is null). Finally, we resume our conclusions in section \ref{conc.}.   

\section{The metric-affine bumblebee model}
\label{sec:1}

We start this section by writing down the metric-affine bumblebee action in a curved space-time (\cite{Delhom:2019gxg}):
\begin{eqnarray}
S_{B}&=&\int d^4 x\,\sqrt{-g}\Big[\frac{1}{2\kappa^2}\Big(R(\Gamma)+\xi B^{\alpha} B^{\beta} R_{\alpha\beta}(\Gamma)\Big)-\frac{1}{4}B^{\mu\nu}B_{\mu\nu}-V(B^{\mu}B_{\mu}\pm b^2)\Big] + \nonumber\\ 
&+& \int d^4 x\,\sqrt{-g}\mathcal{L}_{M}(g_{\mu\nu},\psi),
\label{bumblebee}
\end{eqnarray} 
where $R(\Gamma)$ is the Ricci scalar, $R_{\mu\nu}(\Gamma)$ is the Ricci tensor, $\mathcal{L}_{M}$ is the Lagrangian of matter sources and $B_{\mu}$ is the bumblebee field. This action is defined in the metric-affine approach, which means that the connection is taken to be independent of the metric. { Thus,} one has three degrees of freedom in a way different either from Riemann or from Riemann-Cartan space-time versions \cite{Kostelecky:2003fs}. As usual in LV models, the vector field $B_{\mu}$ acquires a non-zero vacuum expectation value (VEV), say $B_{\mu}=b_{\mu}$,  which in turn is defined as one of the minima of the potential $V$, i.e., $V^{\prime}(B^{\mu}B_{\mu}\pm b^2)=0$, where the prime stands for derivative with respect to the argument of $V$. Under these considerations, observables are treated within a preferred frame in space-time, and the Lorentz symmetry is broken spontaneously. By now, it is worth calling attention we are referring to local Lorentz symmetry breaking since we are dealing with curved spaces. Further, we shall denote $B_{\mu\nu}$ by the bumblebee strength field associated to $B_{\mu}$, and its explicit form is defined below:
\begin{eqnarray}
B_{\mu\nu}= \left(dB\right)_{\mu\nu}.
\end{eqnarray}
Another interesting feature is that the antisymmetric part of the Ricci tensor does not develop a nontrivial contribution in the second term of (\ref{bumblebee}). { Consequently}, it does not contribute to dynamical equations for the torsion.   

\subsection{Field Equations}

To obtain the equations of motion, we must vary the action with respect to the dynamical variables, that is, the metric, the connection, and the bumblebee field. In doing so, we get the following  field equations for $g_{\mu\nu}$, $\Gamma^{\alpha}_{\mu\nu}$, and $B_{\mu}$, respectively:
\begin{align}
	&R_{(\mu\nu)}(\Gamma)-\dfrac{1}{2}g_{\mu\nu}\bigg(R(\Gamma)+\xi B^{\alpha}B^{\beta}R_{\alpha\beta}(\Gamma)\bigg)+2\xi\bigg(B_{(\mu}R_{\nu)\beta}(\Gamma)\bigg)B^{\beta}=\kappa^2  (T^M_{\mu\nu}+T^B_{\mu\nu}),\label{Riccieq}\\
	&{\nabla^{(\Gamma)}_{\lambda}\left[\sqrt{-h} h^{\nu\mu}\right]-\delta^\mu_\lambda \nabla^{(\Gamma)}_{\rho}\left[\sqrt{-h} h^{ \nu\rho}\right]}=\sqrt{-h}\left[{T}^{\mu}_{\lambda \alpha} h^{\nu\alpha}+{T}^{\alpha}_{\alpha \lambda} h^{\nu\mu}-\delta_{\lambda}^{\mu} {T}^{\alpha}_{\alpha \beta} h^{\nu\beta}\right]+\nonumber\\
	&+\kappa^2\Delta_{\lambda}^{\mu \nu},\label{connectioneq}\\
	&\nabla_{\mu}^{(g)}B^{\mu\nu}=-\dfrac{\xi}{\kappa^2}g^{\nu\alpha}B^{\beta}R_{\alpha\beta}(\Gamma)+2 V^{\prime}B^{\nu},\label{bumblebeeeq}
\end{align}
where we have splitted the total stress-energy tensor into two pieces: the first one, coming from contributions of the matter sources $(T_{\mu\nu}^{M})$ and, the second, of the bumblebee field $(T_{\mu\nu}^{B})$. Explicitly, they are defined as follows
\begin{eqnarray}
T_{\mu\nu}^{M}&=&-\frac{2}{\sqrt{-g}}\frac{\delta(\sqrt{-g}\mathcal{L}_{M})}{\delta g^{\mu\nu}},\\
T_{\mu\nu}^{B}&=& B_{\mu\sigma}B_{\nu}^{\ \sigma}-\frac{1}{4}g_{\mu\nu}B^{\alpha}_{\ \sigma}B^{\sigma}_{\ \alpha}-V g_{\mu\nu}+2V^{\prime}B_{\mu}B_{\nu}.
\end{eqnarray}
Also we have defined the object $h^{\mu\nu}$ as
\begin{equation}
\sqrt{-h}h^{\mu\nu}\equiv\sqrt{-g}\big(g^{\mu\nu}+\xi B^{\mu}B^{\nu}\big), \label{definitionh}
\end{equation}
$T^\lambda_{\mu\nu}\equiv2\Gamma^\lambda_{[\mu\nu]}$ is the torsion tensor, and $\Delta_\lambda^{\mu\nu}$ is the hypermomentum describing the coupling between matter and connection at the level of the field equations, which is defined by 
\begin{equation}\label{eq:DefHypermomentum}
\Delta_{\lambda}^{\mu\nu} \equiv 2 \frac{\delta \left(\sqrt{-g}\mathcal{L}_M\right)}{\delta \Gamma^{\lambda}_{\mu\nu}}.
\end{equation}

At first glance, the system of partial differential equations (\ref{Riccieq}-\ref{bumblebeeeq}) seems to be quite complicated to solve since it is coupled. However, it can be done by means of manipulations that { allow} to decouple the equations \cite{Delhom:2019gxg}, so let us see how to proceed further in order to get a more suitable representation of the field equations. First, contracting Eq. (\ref{Riccieq}) with $g^{\mu\nu}$, we obtain the important relation between the Ricci scalar and the trace of the stress-energy tensor ($T=g^{\mu\nu}T_{\mu\nu}$),
\begin{equation}
R(\Gamma)=-\kappa^2 T,\label{Rs}
\end{equation}
{ which has} the same form as in GR. In a similar way, it can be shown that
\begin{eqnarray}
B^{\mu}B^{\nu}R_{\mu\nu}(\Gamma)&=&\frac{\kappa^2}{2+3\xi X}\big(-TX+2B^{\mu}B^{\nu}T_{\mu\nu}\big),\label{Rs2}\\
B^{\mu}R_{\mu\nu}(\Gamma)&=&\frac{\kappa^2}{1+\xi X}\bigg(B^{\mu}T_{\mu\nu}-\frac{\xi B_{\nu}B^{\alpha}B^{\beta}T_{\alpha\beta}+B_{\nu}\left[1+\xi X\right]T}{2+3\xi X}\bigg),\label{Rs3}
\end{eqnarray}
  where $X\equiv g^{\mu\nu}B_{\mu}B_{\nu}$. Putting Eqs. (\ref{Rs}-\ref{Rs3}) into Eq.(\ref{Riccieq}), we arrive at
	\begin{eqnarray}
	R_{(\mu\nu)}(\Gamma)&=&\dfrac{1}{2}\kappa^2 g_{\mu\nu}\bigg[\dfrac{\xi}{2+3\xi X}\bigg(2B^{\alpha}B^{\beta}T_{\alpha\beta}-T X\bigg)-T\bigg]-\dfrac{\xi \kappa^2}{1+\xi X}\bigg[B^{\alpha}\big(B_{\mu}T_{\alpha\nu}+B_{\nu}T_{\alpha\mu}\big)- \nonumber \\
	&-&\dfrac{B_{\mu}B_{\nu}}{2+3\xi X}\bigg(2\big(1+\xi X\big)T+2\xi B^{\alpha}B^{\beta}T_{\alpha\beta}\bigg)\bigg]+\kappa^2 T_{\mu\nu},
	\end{eqnarray}
which means that the Ricci tensor is a function of $B_{\mu}$, $g_{\mu\nu}$, and the matter sources. As a consequence, the non-minimal term $B^{\mu}B^{\nu}R_{\mu\nu}(\Gamma)$ that appears in the action of the model might be rewritten as bumblebee's self-interaction terms, involving couplings between the bumblebee and matter sources and so on.

Now, we turn our attention to the connection equation (\ref{connectioneq}). The structure of the connection field equations in Ricci-based theories without matter couplings to the connection and also allowing minimal coupling has been studied in detail in several works (see as e.g. \cite{Afonso:2017bxr,BeltranJimenez:2017doy}). The most general case for Ricci-based theories is that where nonminimal couplings through the symmetrized Ricci tensor are allowed (as is the case here), and it was studied in full generality in \cite{Delhom:2022vba}. There, it is argued how the connection can be solved as the Levi-Civita connection of the quantity $\sqrt{-g}\partial L/\partial R_{(\mu\nu)}$ where $L$ is the gravitational part of the action and includes the non-minimal couplings of other fields to the symmetrized Ricci tensor and other terms corresponding to hypermomentum contributions.\footnote{Actually, it can be solved in this way up to a projective mode which is physically irrelevant due to the projective symmetry of the action.} One can check that for the theory that we are considering \eqref{bumblebee}, this object is given by the right hand side of \eqref{definitionh} so that, provided that the hypermomentum does not depend on $\Gamma$ as is the case for minimally coupled fermions, the connection in these theories is given by 
\begin{equation}
\Gamma^{\alpha}_{\mu\nu}={}^h\Gamma^\alpha_{\mu\nu}+\Upsilon^\alpha_{\mu\nu},
\label{Chr}
\end{equation}
where 
\begin{equation}
{}^h\Gamma^\alpha_{\mu\nu}=\frac{1}{2}h^{\alpha\lambda}\bigg(-\partial_{\lambda}h_{\mu\nu}+\partial_{\mu}h_{\nu\lambda}+\partial_{\nu}h_{\mu\lambda}\bigg)
\end{equation}
are the Christoffel symbols of the Einstein frame metric $h_{\mu\nu}$ and according to the connection equation \eqref{connectioneq}, $\Upsilon^\lambda_{\mu\nu}$ satisfies
\begin{equation}
\Upsilon^\alpha_{\mu\nu}\lrsq{\delta^\kappa_\alpha\delta^\mu_\beta\delta^\nu_\gamma+\frac{1}{2}\delta^\mu_\alpha\lr{h^{\nu\kappa}h_{\beta\gamma}-\delta^\nu_\beta\delta^\kappa_\gamma-\delta^\nu_\gamma\delta^\kappa_\beta}}=\frac{\kappa^2}{2\sqrt{-h}}h^{\kappa\lambda}\lr{{\Delta}_{\beta\gamma\lambda}+{\Delta}_{\gamma\lambda\beta}+{\Delta}_{\lambda\beta\gamma}}
\label{Upsilon}
\end{equation}
up to a projective mode \cite{Delhom:2022vba},  which is physically irrelevant here since our action has projective symmetry. Let us emphasize that here the indices of the hypermomentum have been lowered using $h_{\mu\nu}$. Thus, wee see that $\Upsilon$ accounts for the hypermomentum contribution, which vanishes for minimally coupled bosonic fields but has a nontrivial contribution for minimally coupled fermionic fields. Its particular form for the case of minimally coupled Dirac fields is derived in appendix \ref{appendix3}. We see that in this case the connection does not propagate any extra degrees of freedom, as it is the case in RBG theories even when they couple nonminimally to the Ricci tensor \cite{Delhom:2022vba}. What occurs in these models is that there is a tricky interplay between the algebraic equations for the metric \eqref{Riccieq} and the connection equations \eqref{connectioneq} (which can both be seen as constraint equations) that leads to non-trivial dynamics for the object $h$, which satisfies Einstein-like field equations, and specifies the connection completely in terms of this object and the matter fields. In order to have new propagating degrees of freedom for the connection, one should include derivative terms of the connection in the action beyond the combinations appearing in the symmetrized Ricci tensor. This can be realized by either including other curvature invariants, or through nonminimal couplings with the matter fields beyond the symmetrized Ricci. A well known example is the case when projective symmetry is broken and the antisymmetric piece of the Ricci is also included in the action. In that case the projective mode acquires dynamics as a massless vector ghost, and the theory ends up propagating an extra ghostly 2-form field which leads to pathological couplings that awake Ostrogradsky instabilities \cite{BeltranJimenez:2019acz,BeltranJimenez:2020sqf} (see also \cite{Delhom:2022vba}). Other examples include gravity actions such as Poincaré gauge gravity or metric-affine gravity with vector kinetic terms where the torsion and nonmetricity tensors can acquire dynamics. While the general case is unstable as well, there are particular subcases which are devoid of instabilities at least around a flat background \cite{BeltranJimenez:2019hrm,Percacci:2020ddy,Jimenez-Cano:2022sds}.

The object $h$ appearing above is defined by \eqref{definitionh}, which in matrix form reads
\begin{equation}
\sqrt{-h}\hat{h}^{-1}=\sqrt{-g}\hat{g}^{-1}\bigg(\hat{I}+\xi \hat{BB}\bigg),
\end{equation}
where $\hat{h}^{-1}$ is matrix notation for $h^{\mu\nu}$ and, consequently, $\hat{h}$ is used for the $h_{\mu\nu}$  matrix (the same notation is valid to $g_{\mu\nu}$ and $g^{\mu\nu}$). Taking the determinant of the last equation, we find
\begin{equation}
\hat{h}=\hat{g}\det{(\hat{I}+\xi \hat{BB})},
\end{equation} 
which, upon insertion into Eq. (\ref{definitionh}), leads to
\begin{equation}
h^{\mu\nu}=\frac{1}{\sqrt{\det{(\hat{I}+\xi \hat{BB})}}}g^{\mu\alpha}(\delta^{\nu}_{\alpha}+\xi B^{\nu}B_{\alpha}).
\label{cch}
\end{equation}
The covariant metric is easily obtained by means  of the relation $h^{\mu\nu}h_{\alpha\nu}=\delta^{\mu}_{\alpha}$. Therefore, 
\begin{equation}
h_{\mu\nu}=\sqrt{\det{(\hat{I}+\xi \hat{BB})}}\bigg[g_{\mu\alpha}\bigg(\delta^{\alpha}_{\nu}-\frac{\xi}{\det{(\hat{I}+\xi \hat{BB})}}B^{\alpha}B_{\nu}\bigg)\bigg].
\label{ch}
\end{equation}

Eqs. (\ref{cch}-\ref{ch}) show how $h_{\mu\nu}$  is related with $g_{\mu\nu}$ and $B_{\mu}$. In the literature, such relations are commonly referred to as disformal transformations \cite{Bek}. In Lorentz-violating scenarios, similar metric structures have been considered even in Riemannian spaces (see, e.g.,~\cite{Seifert:2009gi}). The determinant can be calculated using the method proposed in appendix C  of \cite{Bek}. Here we must note that while the method in \cite{Bek}  can be straightforwardly applied for the cases of time-like or space-like $B_\mu$, it is not so for a light-like $B_\mu$ since $\hat{BB}$ cannot be written in a form required in the proof of \cite{Bek} in any local Lorentz frame for light-like $b_\mu$. Nonetheless, all terms in the expansion in powers of $\xi$ will vanish because the traces of $\hat{BB}$ and the squared matrix $\left(\hat{BB}\right)^2$ are equal to zero for light-like $b_\mu$, and they are the only possible ingredients for building up terms of such expansion. Therefore, for light-like $B_\mu$, the determinant reduces to $1$. As $X=0$ for null $B_\mu$, the formula is valid in any case. The essence of our approach is as follows. We start with the relation
\begin{equation}
\det{(\hat{I}+\xi \hat{BB})}=1+\xi X.
\end{equation} 
In this way, $h_{\mu\nu}$ takes the form
\begin{eqnarray}
h^{\mu\nu}&=&\frac{1}{\sqrt{1+\xi X}}g^{\mu\alpha}(\delta^{\nu}_{\alpha}+\xi B^{\nu}B_{\alpha}),\label{cch1}\\
h_{\mu\nu}&=&\sqrt{1+\xi X}\bigg[g_{\mu\alpha}\bigg(\delta^{\alpha}_{\nu}-\frac{\xi}{1+\xi X}B^{\alpha}B_{\nu}\bigg)\bigg].
\label{ch1}
\end{eqnarray}
Physically speaking, both coefficients (the conformal and disformal ones) depend on the non-minimal Lorentz-violating coupling and the fixed-norm bumblebee field. In addition, it is straightforward to check that by turning off the non-minimal coupling, the effects of the Lorentz symmetry breaking and, equivalently, the non-metricity vanish identically.   

From Eq.~(\ref{Chr}), one finds $\nabla^{(\Gamma)}_{\,\,\mu}h_{\nu\alpha}=0$ which yields
\begin{equation}
Q_{\mu\nu\alpha}=\frac{\xi}{1+\xi X}\left[B_{\alpha}\left(\nabla^{(\Gamma)}_{\,\,\mu}B_{\nu}\right)+B_{\nu}\left(\nabla^{(\Gamma)}_{\,\,\mu}B_{\alpha}\right)\right],
\end{equation}
where the nonmetricity tensor is $\nabla^{(\Gamma)}_{\,\,\mu}g_{\nu\alpha}=Q_{\mu\nu\alpha}$. {As a result}, the non-metricity tensor is specified locally by the bumblebee and its first-order derivatives. 

Let us now turn to the dynamical bumblebee equation. Substituting Eq. (\ref{Rs3}) into Eq. (\ref{bumblebeeeq}), we find that its dependence on the connection is eliminated and, thus, we find a direct relation between $g_{\mu\nu}$ and $b_{\mu}$. The resulting equation looks like the Proca one in the curved space-time looking like
\begin{equation}
\nabla_{\alpha}^{(g)} B^{\alpha\mu}=\mathcal{M}^{\mu}_{\,\,\nu}B^{\nu},
\label{Proca}
\end{equation}
where $\mathcal{M}^{\mu}_{\,\,\nu}$ is the effective mass-squared tensor, and its explicit form is given by
\begin{eqnarray}
\mathcal{M}^{\mu}_{\,\,\nu}=\left(\frac{\xi T}{2+3\xi X}+\frac{\xi^2 B^{\alpha}B^{\beta}T_{\alpha\beta}}{(1+\xi X)(2+3\xi X)}+2V^{\prime}\right)\delta^{\mu}_{\,\,\nu}-\frac{\xi}{(1+\xi X)}T^{\mu\alpha}g_{\nu\alpha}.
\label{masseff}
\end{eqnarray}

Indeed, the equation (\ref{Proca}) exhibits new couplings between the bumblebee field and the energy-momentum tensor displaying no similarities to the metric case. Note that the bumblebee field equation can present instabilities. { In fact}, if the determinant of the mass-squared tensor takes negative values (it is completely possible since the second term has { an opposite sign concerning} the first one in Eq.(\ref{masseff})), { the bumblebee field describes} a tachyonic particle. As a consequence, { in strong-field} regions, the mechanism known as spontaneous tensorization could arise, for example, { near high-density} objects like stars \cite{Ramazanoglu:2017xbl, Ramazanoglu:2019jrr, Cardoso:2020cwo}.

For example, now it is possible to generate a mass term for the background solution. In this situation, the potential vanishes differently from the metric case. Another important issue is the fact that Eq.~(\ref{Proca}) obeys a constraint. In order to check this out, we shall take the divergence of (\ref{Proca}). Its left-hand side vanishes while its right-hand side does not vanish. Therefore, one gets the constraint: 
\begin{equation}
\nabla_{\nu}^{(g)}(m^{2}_{eff}B^{\nu}+J_{eff}^{\nu})=0,
\end{equation}
where
\begin{eqnarray}
	m^{2}_{eff}&=&\frac{\xi T}{2+3\xi X}+\frac{\xi^2 B^{\alpha}B^{\beta}T_{\alpha\beta}}{(1+\xi X)(2+3\xi X)}+2V^{\prime};\nonumber\\
	J_{eff}^{\nu}&=&-\frac{\xi}{(1+\xi X)}T^{\nu\alpha}B_{\alpha}.
\end{eqnarray}

On the other hand, by making the redefinition $\bar{J}_{eff}^{\nu}=m^{2}_{eff}B^{\nu}+J_{eff}^{\nu}$, the constraint equation is interpreted as a conservation law, i.e., as $\nabla_{\nu}^{(g)}\bar{J}_{eff}^{\nu}=0$. 

In the above-mentioned papers, the weak quantum effects of the metric-affine bumblebee gravity in the presence of quantum spinor and scalar matter fields were explored. In other words,  the weak gravity limit $h_{\mu\nu}\approx \eta_{\mu\nu}$ was taken, which means disregarding the Newtonian and post-Newtonian corrections to $g_{\mu\nu}$. It is thus interesting to consider the effects of a nontrivial background geometry. In the next section, we will address this issue by dealing with the effective potential of a fermionic field in the background given by $g_{\mu\nu}$, which, in turn, may be rewritten in terms of $h_{\mu\nu}$ and $B_{\mu}$, making explicit  the dependence on the non-metricity. 
  
\section{Spinor sector and spontaneous Lorentz symmetry breaking}
\label{section3}

In this section, we will consider Dirac spinor fields as the only matter source. In order to implement spinor representations in curved space it is convenient to introduce the vierbein or tetrad formalism. In this approach, the dynamical fields, $g_{\mu\nu}$ and $\Gamma^{\mu}_{\nu\alpha}$ are replaced by new dynamical fields, namely: the tetrad or vierbein field, $e^{a}_{\,\,\mu}$ and the spin connection, $\omega^{ab}_{\,\,\,\,\,\mu}$, respectively.
   
\subsection{Vierbein formalism and fermion action}

The Dirac action in metric-affine theories minimally coupled to the geometry in the sense defined in \cite{Delhom:2020hkb} reads

\begin{equation}
\mbox{S}_{M}=
\int d^{4}x\,\sqrt{-g}\left[\frac{i}{2}e_{\,\,a}^{\mu}\left(\bar{\Psi}\gamma^{a}(\nabla_{\mu}^{(\Gamma)}\Psi)-(\nabla_{\mu}^{(\Gamma)}\bar{\Psi})\gamma^{a}\Psi\right)-m\bar{\Psi}\Psi\right],
\label{ferm}
\end{equation} 
where latin indices are local Lorentz indices running from $0$ to $3$, and we can write the components of tensor fields in spacetime in an anholonomic (non-coordinate) frame through the vierbein\footnote{More technically, latin indices are coordinate fibers in a vector $SO(3,1)$-bundle over spacetime, and the vierbein is the soldering form of such bundle into the tangent bundle which, locally, can be seen as a linear isomorphism and, therefore, is locally invertible. For more technical details on the building of the spinor connection from a general affine connection and on the meaning of the soldering form, see chapter 2 of \cite{Delhom:2022vba} and references therein.} $e^{a}{}_\mu$, so that given such a frame $\left\{e_{a}(x)\right\}$ and its dual frame $\left\{\theta^{a}(x)\right\}$, the metric can be written as
\begin{equation}
g=\eta_{ab}\theta^{a}\otimes\theta^{b}=g_{\mu\nu}dx^{\mu}\otimes dx^{\nu},
\label{gform}
\end{equation}
where $\eta_{ab}$ is the Minkowski metric and the orthonormal (dual) basis is related to a coordinate basis by $\theta^{a}(x)=e^{a}_{\,\,\mu}dx^{\mu}$ and $e_{a}(x)=e_{a}^{\,\,\mu}\partial_{\mu}$. By definition, $e_{a}^{\,\,\mu}(x)$ is the inverse of $e^{a}{}_\mu$and we have $e^{\,\,a}_{\mu}(x) e_{b}^{\,\,\mu}(x)=\delta^{a}_{b}$ and $e^{\,\,\mu}_{a}(x) e_{\ \nu}^{a}(x)=\delta^{\mu}_{\nu}$. The covariant derivative acts on spinors and dual spinors as
 \begin{equation}
 \nabla_{\mu}^{(\Gamma)}\Psi=\partial_{\mu}\Psi+\frac{1}{8}\omega^{(\Gamma)}_{\mu ab}[\gamma^a,\gamma^b]\Psi\quad\text{and}\quad  \nabla_{\mu}^{(\Gamma)}\bar\Psi=\partial_{\mu}\bar\Psi-\frac{1}{8}\omega^{(\Gamma)}_{\mu ab} \bar\Psi[\gamma^a,\gamma^b],
 \label{spinorcovder}
 \end{equation}
where $[\gamma_{a},\gamma_{b}]$ are proportional to the generators of the Lorentz group in the Dirac representation and $\omega^{(\Gamma)}_{\mu ab}$ are the components of the spin connection canonically associated to the affine connection $\Gamma$ through the associated vector bundle construction in a given frame $\{e_a\}$, which are given by
\begin{equation}
\omega^{(\Gamma)}_{\mu ab}=\eta_{ac}e^{c}{}_\nu\left(\partial_\mu e^{\nu}{}_b+e_b{}^\beta \Gamma^\nu_{\mu\beta}\right).
\label{gamu}
\end{equation}
In order to write the Einstein frame version of the above spinor action, we need to define an anholonhomic frame (and dual frame) $\left\{E_{a}(x)\right\}$ (and $\left\{\Theta^{a}(x)\right\}$) as the frame in which the components of $h_{\mu\nu}$ are those of the Minkowski metric in an orthonormal basis, namely
\begin{equation}
h=\eta_{ab} \Theta^{a}\otimes\Theta^{b}=h_{\mu\nu}dx^{\mu}\otimes dx^{\nu}.
\label{hform}
\end{equation}  
Which, again, are (locally) related to coordinate frames by the corresponding vierbein $E^{a}_{\,\,\mu}$ (and its inverse $E_{a}^{\,\,\mu}(x)$) as  $\Theta^{a}=E^{a}_{\,\,\mu}dx^{\mu}$ and $E_{a}=E_{a}^{\,\,\mu}\partial_{\mu}$, respectively. From the relation between the metrics $g$ and $h$ given in \eqref{definitionh}, we can find a relation between the two vierbeins  which reads 
\begin{equation}
\begin{split}
e^{a}{}_{\mu}&=(1+\xi X)^{-1/4}E^{a}_{\,\,\lambda}\left[\delta^\lambda{}_\mu-\frac{(1-\sqrt{1+\xi X})}{X(1+\xi X)^{1/2}}\tilde{B}^{\lambda}B_{\mu}\right];\\
e^{\mu}{}_a&=(1+\xi X)^{1/4}E^\lambda{}_{a}\left[\delta^\mu{}_\lambda+\frac{(1-\sqrt{1+\xi X})}{X(1+\xi X)}\tilde{B}^{\mu}B_{\lambda}\right].
\end{split}
\label{vierrelations}
\end{equation}  
From here on, to denote that an index has been risen or lowered with the Einstein frame metric $h_{\mu\nu}$, or that has been written in the orthonormal frame using $E$ vierbeins, will be denoted by a tilde over the corresponding quantity. From the standard Dirac matrices, defined by $\{\gamma^a,\gamma^b\}=2\eta^{ab}$, one can define what are commonly called as curved space Dirac matrices by using the vierbeins. This leads to define $\gamma^{\mu}=e^{\mu}_{\,\,a}\gamma^{a}$ and $\tilde\gamma^{\mu}=E^{\mu}_{\,\,a}\gamma^{a}$, which satisfy respectively $\left\{\gamma^{\mu},\gamma^{\nu}\right\}=2g^{\mu\nu}$ and $\left\{\tilde{\gamma}^{\mu},\tilde{\gamma}^{\nu}\right\}=2h^{\mu\nu}$. From these relations and definitions, and by splitting the connection as in Eq. \eqref{Chr}, and expanding the covariant derivatives in \eqref{ferm}, one can find the hypermomentum piece due to Dirac fields, which reads (see appendix \ref{appendix3} for details)
\begin{equation}
\begin{split}
\Delta_\lambda^{\mu\nu}=-\frac{\sqrt{-h}}{2}\bigg(E_{\,\,a}^{\mu}+&(1+\xi X)^{1/4}\theta^\mu_a\bigg)\epsilon^{abcd}\ja_d
\lrsq{\eta_{bk}E^k{}_\lambda E^\nu{}_c\right.\\
&\left.+(1+\xi X)^{-1/4}\eta_{bk}E^c{}_\lambda\theta^\nu{}_c-\theta_{b\lambda}\left((1+\xi X)^{1/4} E^\nu{}_c+\theta^\nu{}_c\right)}
\end{split}
\end{equation}
which does not depend on the connection and, therefore provides an algebraic solution to the connection in terms of $h$, $B_\mu$ and the spinor field by means of Eq. $\eqref{Chr}$ with
\begin{eqnarray}
\Upsilon^\kappa_{\beta\gamma}&=&\kappa^2\epsilon^{abcd}\ja_d h^{\kappa\lambda}\Bigg[\lr{h_{\gamma\mu}E_{\,\,a}^{\mu}+\xx^{1/2}\theta_{\gamma a}}\lrsq{\eta_{bk}E^k{}_\beta h_{\lambda\nu} E^\nu{}_c+\xxsm\eta_{bk}E^c{}_\beta\theta_{\lambda c}-\right.\nonumber\\
& & \left. -	\theta_{b\beta}\left(\xxsp h_{\lambda\nu}E^\nu{}_c+\theta_{\lambda c}\right)}\nonumber\\
&&+\lr{h_{\lambda\mu}E_{\,\,a}^{\mu}+\xx^{1/2}\theta_{\lambda a}}\lrsq{\eta_{bk}E^k{}_\gamma h_{\beta\nu} E^\nu{}_c+\xxsm\eta_{bk}E^c{}_\gamma\theta_{\beta c}\right.\nonumber\\ &&\left.
	-\theta_{b\gamma}\left(\xxsp h_{\beta\nu}E^\nu{}_c+\theta_{\beta c}\right)}\nonumber\\
&&+\lr{h_{\beta\mu}E_{\,\,a}^{\mu}+\xx^{1/2}\theta_{\beta a}}\lrsq{\eta_{bk}E^k{}_\lambda h_{\gamma\nu} E^\nu{}_c+\xxsm\eta_{bk}E^c{}_\lambda\theta_{\gamma c}\right.\nonumber\\ && \left.-
	\theta_{b\lambda}\left(\xxsp h_{\gamma\nu}E^\nu{}_c+\theta_{\gamma c}\right)}\Bigg],\label{hypermomexplicit}
\end{eqnarray}
where $\xx=\sqrt{1+\xi X}$. Thus, it is now clear that the connection plays the role of an auxiliary field and does not propagate any new degrees of freedom. Note that the terms which remain in the $\xi\to0$ limit, i.e. those without $\theta$, combine to yield the well known hypermomentum for GR minimally coupled  to a Dirac field, namely $3\kappa^2h^{\kappa\lambda} \epsilon_{\gamma\beta\lambda\rho}\ja^\rho$, which sources the totally antisymmetric piece of the torsion tensor. The other terms contain interactions of the schematic form $\ja B^2$,  $ \xx^{\pm 1/2} \ja B^2$, $\ja B^4$, $\xx\ja  B^4$ and $\xxsp\ja B^6$, where the powers only indicate number of fields of the corresponding species, not contractions.\\

The above solution for the connection shows how the nonminimal coupling of the bumblebee to the Ricci tensor in the action ends up yielding nontrivial nonmetricity and torsion tensors from the point of view of the Einstein frame, as opposed to only totally antisymmetric torsion, as occurs in other theories minimally coupled to fermionic fields. However, the resulting nonmetricity and torsion are just auxiliary fields which can be algebraically solved in terms of the propagating fields of the theory, yielding higher order interactions among them. In this sense, in the result we observe the usual term contributing to the axial piece for the torsion, as well as other contributions due to the presence of the bumblebee field. When this solution for the connection is plugged back into the action, we find that the Dirac action in the Einstein frame of the theory takes the form (see appendix \ref{appendix3} for details)
\begin{equation}
\begin{split}
\mbox{S}_{f}=&\int d^{4}x\,\sqrt{-h}\Bigg\{iE_{\,\,a}^{\mu}\bar{\Psi}\gamma^{a}\nablah_{\mu}\Psi-\frac{m}{(1+\xi X)^{1/4}}\bar{\Psi}\Psi+i\frac{1-\sqrt{1+\xi X}}{X(1+\xi X)^{3/4}}\tilde{B}^\mu \tilde{B}_a\bar{\Psi}\gamma^{a}\nablah_{\mu}\Psi\\
&+i\frac{1-\sqrt{1+\xi X}}{2X(1+\xi X)^{3/4}}\lr{\tilde{T}_\alpha \bar{\Psi}\tilde{\gamma}^\alpha\Psi+i\tilde{\ja}^\alpha \tilde{S}_\alpha}-\kappa^2\Sigma(h,B,\ja)\Bigg\},\label{Sfr}
\end{split}
\end{equation}
where $\tilde{T}_\mu$, $\tilde{S}_\mu$ and $\Sigma$ are defined in appendix \ref{appendix3}. Note that $\Sigma$ stems from the hypermomentum contribution and contains Planck-scale suppressed interaction terms of the schematic form\footnote{The $\ja^2$ term is the well known torsion-induced contact 4-fermion interaction arising in metric-affine GR coupled to minimally coupled fermions \cite{Kibble:1961ba}.} $\ja^2$, $\xx^\alpha \ja^2 B^2$, $\xx^\beta \ja^2 B^4$, $\xx^\gamma \ja^2 B^6$, $\xx^\delta \ja^2 B^8$,$\xx^\epsilon \ja^2 B^{10}$, $\xx \ja^2 B^{12}$ where, again, the powers only indicate number of fields of the corresponding species, and the coefficients can be respectively $\alpha\in\{0,\pm1/2\}$, $\beta\in\{0,\pm1/2,\pm1\}$, $\gamma\in\{0,\pm1/2,1,3/2\}$,$\delta\in\{0,1/2,1,2\}$ and $\epsilon\in\{1/2,3/2\}$. Perturbatively, this means that they can contribute to any (4-fermion + 2n-bumblebee) vertex starting at $n=0$ with a Planck scale suppressed coupling. As a result, these interactions will not be relevant for our analysis of the effective action since they do not contribute to the bumblebee two-point function at the 1-loop level.\\

 We remark that even though the scalar quantity $X=g^{\mu\nu}B_{\mu}B_{\nu}$ is explicitly dependent on $g^{\mu\nu}$, such a dependence can be eliminated by defining a new scalar quantity $Y\equiv h^{\mu\nu}B_{\mu}B_{\nu}=X\sqrt{1+\xi X}$, which in turn can always be inverted for $B_{\mu}$ being a time-like vector, i.e., $X>0$. Therefore, in the above action, $X$ must be interpreted as a function of $Y$, $X=X(Y)$. As far as we know, Eq. (\ref{Sfr}) sets up a novel non-linear fermionic action in the literature. Before we proceed further, it is useful to scrutinize some special cases: the first one corresponds to taking the limit $\xi\rightarrow 0$, when the standard fermionic action in metric-affine GR is recovered; and the gravitational and gauge sector reduce, as expected, to the Einstein action plus kinetic and potential terms for the bumblebee, respectively \cite{Kostelecky:2003fs}. As well, in the weak field limit, $h_{\mu\nu}\approx\eta_{\mu\nu}$, and for a perturbative coupling $\xi\ll1$, the above expressions consistently recover the results previously found in \cite{Delhom:2019gxg}.

For our next purpose, it will be convenient to rewrite Eq. (\ref{Sfr}) using compact notation as
\begin{equation}
\mbox{S}_{f}=\int d^{4}x\sqrt{-h}\Bigg\{\bar{\Psi}\left(i\tilde{\Gamma}^{\mu}\tilde{\nabla}_{\mu}-M\right)\Psi-\kappa^2\Sigma(h,B,\ja)\Bigg\}
\label{effA}
\end{equation}
where $\tilde{\nabla}\equiv \nabla^{h}$ and the operators $\Gamma^{\mu}$ and $M$ are expanded in the basis of 16 Dirac matrices in the spinor space of the Clifford algebra. Explicitly, 
\begin{equation}
\begin{split}
\tilde{\Gamma}^{\mu}&=\tilde{\gamma}^{\mu}+c^{\mu}_{\alpha}\tilde{\gamma}^{\alpha}+d^{\mu}_{\alpha}\gamma_{5}\tilde{\gamma}^{\alpha}+e^{\mu} {I}+i f^{\mu}\gamma_{5}+\frac{1}{2}g^{\mu\lambda\alpha}\sigma_{\lambda\alpha},\\
M&= m_{0}+im_{5}\gamma_{5}+a_{\mu}\tilde{\gamma}^{\mu}+k_{\mu}\tilde{\gamma}^{\mu}\gamma_{5}+\frac{1}{2}H^{\mu\nu}\sigma_{\mu\nu},
\end{split}
\end{equation} 
where $c^{\mu}_{\alpha}$, $d^{\mu}_{\alpha}$, $e^{\mu}$, $f^{\mu}$, $g^{\mu\lambda\alpha}$, $m_{5}$, $a_{\mu}$, $k_{\mu}$ and $H^{\mu\nu}$ are Lorentz- and/or CPT-dynamical violating coefficients and we have defined $m_{0}=\frac{m}{(1+\xi X)^{1/4}}$. As it turns out from Eq.(\ref{Sfr}), the only non-zero coefficients are completely given in terms of the bumblebee field, and read
\begin{equation}
\begin{split}
c^{\mu}_{\alpha}&=\xi^{(1)}B_{\alpha}\tilde{B}^{\mu};\\
a_{\mu}&=i\xi^{(2)}\tilde{T}_{\mu};\\
k_{\mu}&=-\xi^{(2)}\tilde{S}_{\mu},
\label{cak}
\end{split}
\end{equation}
with
\begin{equation}
\begin{split}
\xi^{(1)}&=\xi^{(1)}(\xi,X)=\frac{(1-\sqrt{1+\xi X)}}{X(1+\xi X)};\\
\xi^{(2)}&=\xi^{(2)}(\xi,X)=\frac{(1-\sqrt{1+\xi X})}{2X(1+\xi X)^{3/4}}.
\label{xi1}
\end{split}
\end{equation}

The constant couplings $\xi^{(i)}$ previously defined are non-linearly dependent on $\xi$ and $X$. In particular, as long as we restrict to small $\xi$ situation, the $\xi^{(i)}$ no longer depend on $X$, but only linearly on $\xi$. Explicitly, { for} $\xi^{(1)}=-\frac{\xi}{2}$ and $\xi^{(2)}=-\frac{\xi}{4}$, { one} recovers the results found in \cite{Delhom:2019gxg}, as mentioned before.

\subsection{Lorentz-violating coefficients in an effective Minkowskian theory}

In order to clarify the link between the spontaneous Lorentz symmetry breaking and sources of the non-metricity in our model, we shall focus our attention only on these two subjects. To do so, we will neglect the Lorentz-violating contributions stemming from the other sources. {Therefore, in the Einstein frame, we disregard the gravitation effects that are} reasonably attainable around the Earth's surface, where the Newtonian and post-Newtonian corrections can be neglected. In this situation, the vierbein is approximated to $E^{a}_{\,\mu}\approx \delta^{a}_{\,\mu}$. Consequently, the spin connection is $\omega^{(h)}_{ab\mu}\approx 0$.

 Within this scenario, the most interesting physical situation comes up by taking fluctuations around the bumblebee VEV, i.e., 
\begin{equation}
B_{\mu}=b_{\mu}+A_{\mu},
\label{Bmu}
\end{equation}
 where $A_{\mu}$ plays the role of fluctuations.	As a result, the dynamical coefficients for Lorentz violation defined in the former subsection might be also 
expanded around their VEV's in the following way:
\begin{equation}
\begin{split}
c^{\mu}_{\alpha}&=\bar{c}^{\mu}_{\alpha}+C^{\mu}_{\alpha},\\
a_{\mu}&=\bar{a}_{\mu}+D_{\mu},\\
k_{\mu}&=\bar{k}_{\mu}+K_{\mu},
\label{CFF}
\end{split}
\end{equation}
where $\bar{c}^{\mu}_{\alpha}, \bar{a}_{\mu}$ and $\bar{k}_{\mu}$ stand for the VEV's of $c^{\mu}_{\alpha}, a_{\mu}$ and $k_{\mu}$, respectively. Their fluctuations are described by $C^{\mu}_{\alpha}, D_{\mu}$ and $K_{\mu}$. Substituting Eq.~(\ref{Bmu}) in Eq.~(\ref{cak}) and then comparing with Eq. (\ref{CFF}), one makes the following identifications:
\begin{equation}
\begin{split}
\bar{c}^{\mu}_{\alpha}&=\xi^{(1)}b_{\alpha}b^{\mu},\\
\bar{a}_{\mu}&=0,\\
\bar{k}_{\mu}&=0,\\
C^{\mu}_{\alpha}&=\xi^{(1)}\left(b_{\alpha}A^{\mu}+A_{\alpha}b^{\mu}\right),\\
D_{\mu}&=i\xi^{(3)}\left(b_{\mu}\partial_{\nu}A^{\nu}+b^{\nu}\partial_{\nu}A_{\mu}\right),\\
K_{\mu}&=\xi^{(4)}b^{\alpha}F^{\beta\sigma}\epsilon_{\alpha\beta\sigma\mu},
\label{coff}
\end{split}
\end{equation}
where we have defined $\xi^{(3)}=\frac{(1-\sqrt{1+\xi X})}{2X(1+\xi X)}$, $\xi^{(4)}=\frac{(1-\sqrt{1+\xi X})^2}{8X(1+\xi X)^{5/4}}$ and $F^{\beta\sigma}=\partial^{\beta}A^{\sigma}-\partial^{\sigma}A^{\beta}$. The LV coefficient $\bar{c}^{\mu}_{\alpha}$ depends on the fluctuations since the coupling constant $\xi^{(1)}$ is explicitly {dependent on the scalar $X$}, as it can be checked from Eq. (\ref{xi1}). On the other hand, as pointed out before, by assuming $\xi$ to be small, the coupling constant $\xi^{(1)}$  turns out to be $X$-independent at the leading order. Then, $\bar{c}^{\mu}_{\alpha}$ is completely given in terms of the bumblebee VEV and can be interpreted as an effective coefficient $(c_{\mbox{eff}})^{\mu}_{\alpha}$ similarly to SME \cite{Kostelecky:2003fs}. The other coefficients mix the VEV with fluctuations of the bumblebee field.

Making use of the above definitions, we are able to find the quadratic fermionic action in the post-Minkowskian limit. Hence, it looks like  
 \begin{equation}
\begin{split}
\tilde{S}_{f}&=\int d^{4}x \bar{\Psi}\left(i\gamma^{\mu}\partial_{\mu}+i\xi^{(1)}b_{\alpha}b^{\mu}\gamma^{\alpha}\partial_{\mu}-m_{0}\right)\Psi+S^{\prime}\\
&=\int d^{4}x \bar{\Psi}\left(i\gamma^{\mu}\partial_{\mu}+i \bar{c}^{\mu}_{\alpha}\gamma^{\alpha}\partial_{\mu}-m_{0}\right)\Psi+S^{\prime},
\label{new1}
\end{split}
\end{equation}
where $S^{\prime}$ represents the interaction terms among the background, the fluctuation, and the fermionic fields. We find that in this case, an aether-like CPT-even term for the spinor field \cite{CarTam} emerges.

Now, let us remark some comments about this action drawing a parallel with SME. The first two terms can be rearranged by defining an effective fermionic metric, namely, $\eta^{\mu\nu}_{\mbox{eff}}=\eta^{\mu\nu}+\xi^{(1)}b^{\mu}b^{\nu}$. The second part of the effective metric encodes information on the non-metricity in terms of the conformal factor by means of $\xi^{(1)}$ and the disformal part represented by $b_{\mu}b_{\nu}$. One can rewrite the action (\ref{new1}) in terms of the new metric as follows:
\begin{equation}
\tilde{S}_{f}=\int d^{4}x \bar{\Psi}\left(i\eta^{\mu\alpha}_{\mbox{eff}}\gamma_{\alpha}\partial_{\mu}-m_{0}\right)\Psi+S^{\prime}.
\end{equation}  

At the classical level, the dynamical properties of fermions change due to local corrections coming from the bumblebee VEV. For example, in this theory, fermions propagate with an effective mass $m_{0}$ and follow geodesics of the effective metric, $\eta^{\mu\nu}_{\mbox{eff}}$\footnote{Similar results have been found in the metric bumblebee model context \cite{Seifert:2009gi, Bluhm:2004ep, Bluhm:2007bd}, and more general vector-tensor models \cite{Seifert:2009vr, Bailey:2006fd}}, instead of $\eta^{\mu\nu}$. Such modifications could potentially trigger instabilities. For example, if $\xi X<-1$, then one gets a complex effective mass, generating  in this way a tachyonic-like instability. Ghost-like instabilities can also arise. For a discussion of the instabilities of this model at the perturbative level in $\xi$ the reader is referred to \cite{Delhom:2019gxg}. In order to avoid undesired instabilities in the effective theory, one can impose that the coupling constants controlling the LV coefficients should be Planck-suppressed. { In other words,} if $\xi^{(i)}$ are Planck-suppressed,  instabilities would only become important at the Planck scale, where the validity of the effective theory breaks down, and then a full theory should be taken into account.      

Regarding the contributions of the fluctuations, we have observed the emergence of { nonminimal couplings terms (see f.e. \cite{HDCS})}. To see that, we can explicitly write $S^{\prime}$ to get
\begin{equation}
\begin{split}
S^{\prime}=\int d^{4}x\left(D_{\mu}\bar{\Psi}\gamma^{\mu}\Psi+K_{\mu}\bar{\Psi}\gamma^{\mu}\gamma_{5}\Psi\right),
\end{split}
\end{equation}
with $D_{\mu}$ and $K_{\mu}$ defined in Eq.~(\ref{coff}). {Besides being a LV coefficient, $K_{\mu}$ is also a CPT-violating one}. Formally, it behaves as a torsion-like term  in comparison to SME. In fact, even though { there are} similarities with torsion, such an effect is thoroughly due to the sources of  non-metricity. As far as we know, it sets up the first example of LV and CPT-violating term entirely induced by the source of non-metricity,  differently from \cite{Foster:2016uui}, where the non-metricity is assumed to be an external background field coupled to fermions. Secondly, $K_{\mu}$ is an axial coefficient as explicitly displayed in (\ref{coff}) coupled to the fermion axial current defined by $J^{\mu}_{5}=\bar{\Psi}\gamma^{\mu}\gamma_{5}\Psi$. Thus, an axial-like term is generated by non-trivial effects coming from the VEV of sources of nonmetricity.  Within the context of radiative corrections,  this subject has been extensively discussed in the literature (see f.e. \cite{ourrev} and references therein). 

At the perturbative level of the non-minimal coupling, $\xi\ll1$, $K_{\mu}$ does not contribute at the leading order; as a result, one recovers the results found in \cite{Delhom:2019gxg}. Moreover, { we assume that the VEV has no dynamics. In that case, the corresponding LV coefficient disappears even in the {non-perturbative} regime, making it clear that the effective Minkowskian theory is non-trivial only when a non-constant VEV is considered. Otherwise, any Lorentz-violating terms can arise in the fermion sector. 
   
\section{One-loop corrections to the spinor effective action}
\label{section4}

\subsection{Spinor effective action}

In this section, we will calculate the one-loop effective action for the model given by Eq. (\ref{effA}). We start { by using} the background field method, which corresponds to expanding the  classical spinor field around the given background. We assume that gravity is not quantized, i.e., all fields describing the gravitational sector are assumed to be purely background ones. So, setting $\bar{\Psi} \rightarrow \bar{\Psi}+\sqrt{\hslash}\bar{\psi}$ and $\Psi \rightarrow \Psi+\sqrt{\hslash}\psi$, where $\bar{\psi}$ and $\psi$ are quantum fields representing fluctuations around their background fields. So, the spinor effective action looks like
\begin{equation}
e^{i\frac{\Gamma[\bar{\Psi},\Psi]}{\hslash}}=\int \mathcal{D}\bar{\psi}\mathcal{D}\psi\,e^{\frac{i}{\hslash}\left\{S[\bar{\Psi}+\sqrt{\hslash}\,\bar{\psi}, \Psi+\sqrt{\hslash}\,\psi]+\sqrt{h}(\bar{\eta}\,\psi+\bar{\psi}\,\eta)\right\}},
\end{equation}
where we used the shorthand notation: $\bar{\eta}\,\psi\equiv\int d^{4}x\sqrt{-h}\,\bar{\eta}(x)\psi(x)$ and\\ $\bar{\psi}\,\eta\equiv\int d^{4}x\sqrt{-h}\,\bar{\psi}(x)\eta(x)$. Here $\bar{\eta}(x)$ and $\eta(x)$ are sources for the spinor fields $\psi(x)$ and $\bar{\psi}(x)$, respectively. In general, the former equation is fairly complicated to be integrated over the quantum fields. The standard procedure to circumvent this difficulty consists in computing it perturbatively, that is, to expand the effective action in power series of $\hslash$ ({that is, to obtain the} loop expansion), 
\begin{equation}
\Gamma[\bar{\Psi},\Psi]=S[\bar{\Psi},\Psi]+\hslash\,\Gamma^{(1)}[\bar{\Psi},\Psi]+ \mathcal{O}(\hslash^{2}), 
\end{equation}
where $\Gamma^{(1)}$ is the one-loop level effective action. Accordingly, we will restrict our analysis up to the one-loop level and in the vacuum sector, i.e., assuming a vanishing background spinor field. Thus, the spinor contribution to the one-loop effective action is 
\begin{equation}
\Gamma^{(1)}[h_{\mu\nu},B_{\mu}]=-i\ln \int \mathcal{D}\bar{\psi}\mathcal{D}\psi\,e^{i\int d^{4}x\sqrt{-h}\,\bar{\psi}(x)\Delta\psi(x)},
\label{gamma1}
\end{equation} 
where 
\begin{eqnarray}
\nonumber\Delta[h_{\mu\nu},B_{\mu}]&=&i\tilde{\Gamma}^{\mu}\tilde{\nabla}_{\mu}-M\\
\nonumber&=&i\tilde{\gamma}^{\mu}\tilde{\nabla}_{\mu}+i\frac{(1-\sqrt{1+\xi X})}{X(1+\xi X)}(\tilde{\gamma}^{\alpha}B_{\alpha})\tilde{B}^{\mu}\tilde{\nabla}_{\mu}-i\frac{(1-\sqrt{1+\xi X})}{X(1+\xi X)^{3/4}}\tilde{\gamma}^{\mu}\tilde{T}_{\mu}+\\
&+&\frac{(1-\sqrt{1+\xi X})}{2 X(1+\xi X)^{3/4}}\tilde{\gamma}^{\mu}\gamma_{5}\tilde{S}_{\mu}-\frac{m}{(1+\xi X)^{1/4}},
\end{eqnarray} 
{i.e., $\Delta$ is a Dirac operator}. Usually, Eq.(\ref{gamma1}) can be set in a functional determinant form, {given by}
\begin{equation}
\label{det1}\Gamma^{(1)}[h_{\mu\nu},B_{\mu}]=-i\ln\det\Delta=-i\mbox{Tr}\ln\Delta.
\end{equation} 

Since the above equation is divergent, a regularization procedure ought to be adopted in order to calculate the one-loop divergent contributions to { the fermionic effective action}. In the next section, we will provide a general expression for the one-loop divergences by using the Schwinger-DeWitt proper-time method.
  
\subsection{One-loop divergence contributions to the fermionic effective action}

In order to evaluate the one-loop effective action, we will use the  Barvinsky-Vilkovisky technique \cite{Barvinsky:1985an}, along with the dimensional regularization procedure. Actually, the method is an elegant technique to find one-loop divergences for operators in curved spaces with non-minimal couplings, which is our situation as we will see next. Besides that, the method guarantees general covariance throughout the calculations.

We now rewrite the kernel of the functional determinant Eq.~(\ref{det1}) in terms of the operator of second order in derivatives, as usual, to  apply the standard methodology for finding the one-loop divergences in the spinor sector, see for example  \cite{DeBerredoPeixoto:2001qm, Netto:2014faa, Buchbinder:2017zaa}. The trick consists on multiplying the $\Delta$ operator by $\gamma^{2}_{5}=I$ and moving one of the $\gamma_5$ matrices through $\Delta$ in order to use the properties of Dirac gamma matrices to obtain 
\begin{equation}
\mbox{Tr}\ln\Delta=\mbox{Tr}\ln\Delta^{*},
\end{equation}
where 
\begin{eqnarray}
\nonumber\Delta^{*}
&=&i\tilde{\gamma}^{\mu}\tilde{\nabla}_{\mu}+i\frac{(1-\sqrt{1+\xi X})}{X(1+\xi X)}(\tilde{\gamma}^{\alpha}B_{\alpha})\tilde{B}^{\mu}\nabla_{\mu}^{(h)}-i\frac{(1-\sqrt{1+\xi X})}{X(1+\xi X)^{3/4}}\tilde{\gamma}^{\mu}\tilde{T}_{\mu}+\\
&+&\frac{(1-\sqrt{1+\xi X})}{2 X(1+\xi X)^{3/4}}\tilde{\gamma}^{\mu}\gamma_{5}\tilde{S}_{\mu}+\frac{m}{(1+\xi X)^{1/4}},\nonumber
\end{eqnarray}
 which formally can be written as $\Delta^*\equiv i\tilde{\Gamma}^{\mu}\tilde{\nabla}_{\mu}+M^{*}$.
Having this in mind,  we can write down the one-loop effective action as a trace of the logarithm of the second-derivative operator, as follows:
\begin{equation}
\Gamma^{(1)}[h_{\mu\nu},B_{\mu}]=-\frac{i}{2}\ln\det(\hat{\Delta}\cdot\hat{\Delta}^{*})=-\frac{i}{2}\mbox{Tr}\ln(\hat{\Delta}\cdot\hat{\Delta}^{*}).
\label{Qac}
\end{equation}

In order to proceed further we shall focus our attention on the weak non-minimal coupling regime, i.e., $\xi\ll1$. Such a condition is reasonable because Lorentz-breaking effects are assumed to be suppressed by a high energy scale \cite{Kostelecky:2000mm}. Taking this into account, the one-loop effective action is simplified drastically. Thus, carrying out the above product, we have
\begin{equation}
\begin{split}
\hat{H}&=\hat{\Delta}\cdot\hat{\Delta}^{*}=\left(i\tilde{\Gamma}^{\mu}\tilde{\nabla}_{\mu}-M\right)\left(i\tilde{\Gamma}^{\mu}\tilde{\nabla}_{\mu}+M^{*}\right)\\
&=-\tilde{\Gamma}^{\mu}\tilde{\Gamma}^{\nu}\tilde{\nabla}_{\mu}\tilde{\nabla}_{\nu}-\tilde{\Gamma}^{\mu}(\tilde{\nabla}_{\mu}\tilde{\Gamma}^{\nu})\tilde{\nabla}_{\nu}+i\tilde{\Gamma}^{\mu}\tilde{\nabla}_{\mu}M^{*}+i\tilde{\Gamma}^{\mu}M^{*}\tilde{\nabla}_{\mu}-iM\tilde{\Gamma}^{\mu}\tilde{\nabla}_{\mu}-\\
&-MM^{*}.
\end{split}
\end{equation} 
After an exhaustive algebraic manipulation we can put it into the following form:
\begin{equation}\label{eq:H_dWSch}
\hat{H}=-\left(\tilde{\square}\hat{1}+\hat{H}^{\mu\nu}\tilde{\nabla}_{\mu}\tilde{\nabla}_{\nu}+2\hat{X}^{\mu}\tilde{\nabla}_{\mu}+\hat{\Pi}\right),
\end{equation}
{with}
\begin{equation}
\begin{split}
\tilde{\square}&=h^{\mu\nu}\tilde{\nabla}_{\mu}\tilde{\nabla}_{\nu},\\
\hat{H}^{\mu\nu}&=-\xi B^{\mu}B^{\nu}\hat{1}+\mathcal{O}(\xi^2),\\
\hat{X}^{\mu}&=\frac{1}{4}\xi \tilde{T}^{\mu}\hat{1}-\frac{1}{4}\xi \tilde{\sigma}^{\mu\nu}\gamma_{5}\tilde{S}_{\nu}-\frac{1}{4}\xi \tilde{\gamma}^{\nu}\tilde{\gamma}^{\beta}\tilde{\nabla}_{\nu}(B_{\beta}\tilde{B}^{\mu})+\mathcal{O}(\xi^2),\\
\hat{\Pi}&=\frac{1}{4}\xi \left(\tilde{\nabla}^{\mu}\tilde{T}_{\mu}\right)\hat{1}+\frac{1}{8}\xi\tilde{\gamma}^{\mu}\tilde{\gamma}^{\nu}\tilde{F}_{\mu\nu}+\frac{i}{4}\xi \gamma_{5}\tilde{\nabla}^{\mu}\tilde{S}_{\mu}+\frac{i}{8}\xi\tilde{\gamma}^{\mu}\tilde{\gamma}^{\nu}\gamma_{5}\tilde{S}_{\mu\nu}-\frac{1}{4}\tilde{R}\hat{1}+m_{0}^{2}\hat{1}+\\
&+\xi B^{\nu}B_{\alpha}\tilde{R}_{\nu\mu}\tilde{\gamma}^{\alpha}\tilde{\gamma}^{\mu}+\mathcal{O}(\xi^2),
\label{def2}
\end{split}
\end{equation}
and where we have defined 
\begin{equation}
\begin{split}
\tilde{F}_{\mu\nu}=\tilde{\nabla}_{\mu}\tilde{T}_{\nu}-\tilde{\nabla}_{\nu}\tilde{T}_{\mu}, \,\,\, \tilde{S}_{\mu\nu}=\tilde{\nabla}_{\mu}\tilde{S}_{\nu}-\tilde{\nabla}_{\nu}\tilde{S}_{\mu}.
\label{def3}
\end{split}
\end{equation}
Recalling that $\tilde{T}_{\mu}$ and $\tilde{S}_{\mu}$ must be expanded up to first order in $\xi$ in Eqs. (\ref{def2}) and (\ref{def3}).  

Strictly speaking, the presence of the non-minimal term $\hat{H}^{\mu\nu}$ spoils the direct applicability of the standard Schwinger-DeWitt method \cite{Buchbinder:1992rb}, though the calculation can be done by means of the method developed by Barvinsky and Vilkovisky \cite{Barvinsky:1985an}.  As argued  there, the method only works if the non-minimal term is defined in terms of a continuous parameter. Our case fits in this situation since the non-minimality is parametrized by the continuous small parameter $\xi$. In \cite{Netto:2014faa, Buchbinder:2017zaa}, the authors have applied this technique to calculate the divergences in the gauge sector. We will proceed in a similar way. Therefore, the first step is to separate the non-minimal ($\hat{H}_{nm}$) and minimal ($\hat{H}_{m}$) pieces of the Dirac operator, explicitly, 
\begin{equation}
\begin{split}
\hat{H}_{m}&=-\left(\tilde{\square}\hat{1}+2\hat{X}^{\mu}\tilde{\nabla}_{\mu}+\hat{\Pi}\right);\\
\hat{H}_{nm}&=-\hat{H}^{\mu\nu}\tilde{\nabla}_{\mu}\tilde{\nabla}_{\nu}.
\end{split}
\end{equation}
By doing so, one can rewrite Eq. (\ref{Qac}) as follows
\begin{equation}
\begin{split}
\Gamma^{(1)}&=-\frac{i}{2}\mbox{Tr}\ln \hat{H}=\frac{i}{2}\mbox{Tr}\ln (\hat{H}_{m}+\hat{H}_{nm})=-\frac{i}{2}\mbox{Tr}\ln \hat{H}_{m}-\frac{i}{2}\mbox{Tr}\ln(\hat{1} +\hat{H}^{-1}_{m}\hat{H}_{nm})\\
&=\Gamma^{(1)}_{m}-\frac{i}{2}\mbox{Tr}(\hat{H}_{nm}\hat{H}^{-1}_{0})+...,
\label{effac}
\end{split}
\end{equation}    
where  $\Gamma^{(1)}_{m}$ is the purely minimal part of the one-loop effective action, the ellipsis stands for higher-order terms in $\xi$, and $H^{-1}_{0}$ is the inverse of $H_{0}\equiv \tilde{\square}-\frac{1}{4}\tilde{R}$. We only expand the logarithmic term up to the first order in $\xi$ in the last line of the former equation. The explicit calculation of the divergent piece of the contributions coming from the non-minimal one-loop action is shown in Appendix \ref{appendix33}.

 The minimal contributions might be computed directly from the Schwinger-DeWitt method which consists on expanding  $\hat{H}_{m}$ in a power series of the heat kernel coefficient, $a_{j}(x,x^{\prime})$ \cite{heatk}. Using the Green's function proper-time representation, we have
\begin{equation}
\hat{G}(x,x^{\prime})=i\int_{0}^{\infty} ds\, e^{is \hat{H}_{m}}\delta(x,x^{\prime}).
\end{equation} 
Then, the proper-time representation for the one-loop effective action is
\begin{equation}
\Gamma^{(1)}_{m}=-\frac{i}{2}\mbox{Tr}\int_{0}^{\infty}\frac{ds}{s}e^{is \hat{H}_m}.
\end{equation} 
The above operator has a well-known integral representation that results in 
\begin{equation}
\Gamma^{(1)}_{m}=-\frac{i}{2}\int_{0}^{\infty} \frac{i ds}{s}\frac{\mathcal{D}^{\frac{1}{2}}(x,x^{\prime})}{(4\pi is)^{D/2}}e^{(-is m_{0}^2+\frac{i}{2s}\sigma(x,x^{\prime}))}\sum_{j=0}^{\infty} a_{j}(x,x^{\prime})(is)^{j},
\label{divo}
\end{equation}
where $D=4+\epsilon$, with $\epsilon$ being the parameter of dimensional regularization. In addition,
\begin{equation}
\begin{split}
\sigma(x,x^{\prime})&=\frac{1}{2}\tilde{\nabla}^{\mu}\sigma\tilde{\nabla}_{\mu}\sigma,\\
\mathcal{D}(x,x^{\prime})&=\left|\mbox{det}\left(\frac{\partial^{2}\sigma}{\partial x^{\mu}\partial x^{\prime \nu}}\right)\right|,
\end{split}
\end{equation}
are the geodesic distance and the Van Vleck-Morette determinant, respectively. We are following the definitions of \cite{heatk1}. 

In this approach the divergent part of the one-loop effective action for the operator of the general form given by Eq. (\ref{divo}) is 
\begin{equation}
\Gamma^{(1)}_{m}\Big|_{\mbox{div}}=\frac{\mu^{D-4}}{(4\pi)^{2}\epsilon}\int d^{D}x\sqrt{-h}\,\mbox{tr}\left(\lim_{x\rightarrow x^{\prime}}a_{2}(x,x^{\prime})\right),
\label{hhhh}
\end{equation}
with $\mu$ being the mass scale parameter introduced within the framework { of the dimensional regularization}. The second-order heat kernel coefficient is given by 
\begin{equation}
\begin{split}
\lim_{x\rightarrow x^{\prime}}a_{2}(x,x^{\prime})&=\frac{1}{180}\left(\tilde{R}_{\mu\nu\alpha\beta}\tilde{R}^{\mu\nu\alpha\beta}-\tilde{R}_{\mu\nu}\tilde{R}^{\mu\nu}+\tilde{\square}\tilde{R}\right)\hat{1}+\frac{1}{2}\hat{P}^2+\\
&+\frac{1}{12}\tilde{W}^{\mu\nu}\tilde{W}_{\mu\nu}+\frac{1}{6}\tilde{\square}\hat{P},
\label{a2}
\end{split}
\end{equation}
where we have the following definitions
\begin{eqnarray}
\label{p1}\hat{P}&=&\hat{\Pi}+\frac{\tilde{R}}{6}\hat{1}-\tilde{\nabla}_{\mu}\hat{X}^{\mu}-\hat{X}_{\mu}\hat{X}^{\mu},\\
\label{w1}W_{\mu\nu}&=&[\tilde{\nabla}_{\nu},\tilde{\nabla}_{\mu}]\hat{1}+2\tilde{\nabla}_{[\nu}\hat{X}_{\mu]}+2\hat{X}_{[\nu}\hat{X}_{\mu]}.
\end{eqnarray}
Using Eq.(\ref{a2}) (see Appendix \ref{appendix2})  and plugging Eqs. (\ref{first}-\ref{last}) into Eq. (\ref{nm}), we obtain the full one-loop divergent effective action 
\begin{equation}
\begin{split}
\Gamma^{(1)}\Big|_{\mbox{div}}&=\frac{\mu^{D-4}}{(4\pi)^{2}\epsilon}\int d^{D}x\sqrt{-h}\Bigg\{\left(1-\frac{\xi X}{4}\right)\left[\frac{1}{72}\tilde{R}^{2}-\frac{7}{360}\tilde{R}_{\mu\nu\alpha\beta}\tilde{R}^{\mu\nu\alpha\beta}-\frac{1}{45}\tilde{R}_{\mu\nu}\tilde{R}^{\mu\nu}\right]+\\
&+2m^{4}-\frac{1}{3}m^{2}\tilde{R}+\frac{1}{12}\xi\tilde{R}\tilde{\nabla}_{\mu}\tilde{\nabla}_{\nu}\left(B^{\mu}B^{\nu}\right)-m^{2}_{0}\xi\tilde{\nabla}_{\mu}\tilde{\nabla}_{\nu}\left(B^{\mu}B^{\nu}\right)-2m^{4}\xi X+\\
&+4m^{2}\xi B^{\mu}B^{\nu}\tilde{R}_{\mu\nu}-\frac{4}{9}\xi B^{\mu}B^{\nu}\tilde{R}\tilde{R}_{\mu\nu}+\frac{1}{12}\xi\tilde{R}^{\mu\nu\theta\sigma}\tilde{\nabla}_{\mu}\tilde{\nabla}_{\sigma}\left(B_{\theta}B_{\nu}\right)+\\
&-\frac{7}{30}\xi B^{\mu}B^{\nu}\tilde{\nabla}_{\mu}\tilde{\nabla}_{\nu}\tilde{R}-\frac{2}{45}\xi B^{\mu}B^{\nu}\tilde{R}_{\mu\alpha}\tilde{R}^{\alpha}_{\,\nu}+\frac{1}{30}\xi B^{\mu}B^{\nu}\tilde{\square}\tilde{R}_{\mu\nu}-\frac{1}{6}\xi X\tilde{R}^2+\\
&+\frac{1}{20}\xi X\tilde{\square}\tilde{R}+\frac{4}{45}\xi B^{\mu}B^{\nu}\tilde{R}^{\alpha\beta}\tilde{R}_{\alpha\mu\beta\nu}
\Bigg\}.
\label{434}
\end{split}
\end{equation}
where the boundary terms have been thrown away. As expected, this theory presents the same renormalizability issues as GR \cite{Shapiro:2001rz},  which we did not intend to solve here in any way. For this reason, we will restrict ourselves to the Minkowiskian effective theory. In this case, the pure curvature divergences vanish throughout and only two terms remain, namely,
\begin{equation}
\label{Gd2}
\Gamma^{(1)}_{\mbox{div}}=\frac{\mu^{D-4}}{(4\pi)^{2}\epsilon}\int d^{D}x\left[2m^{4}-2m^{4}\xi X+\mathcal{O}(\xi^2)\right].
\end{equation}     
The first term contributes to the vacuum energy corresponding to a bubble Feynman diagram and, as usual, it can be ignored. Regarding the second term, it corresponds to the diagram with $B_{\mu}$ external legs displayed in Fig.\ref{fig:fig1}, a result already obtained in \cite{pj20} by using the diagrammatic method. 
\begin{figure}
	\centering
		\includegraphics[width=0.50\textwidth]{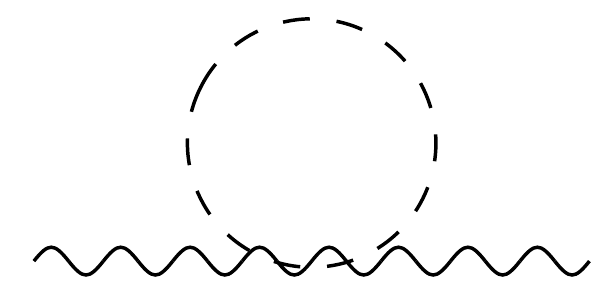}
	\caption{Contribution to the spinor one-loop level.}
	\label{fig:fig1}
\end{figure}

\section{Summary and conclusion}
\label{conc.}

We have studied the metric-affine formulation of bumblebee gravity coupled to a fermion field representing the matter sector. We have concluded that as a result of the affinity properties (non-metricity) of space-time, the theory can be rewritten in the Einstein frame, which seems to be the natural one to explore LSB issues in the matter sector of the model. As we have seen, the spinor action in this frame acquires new (non-minimal) couplings between the bumblebee and the fermion field, which can be cast into the general structure proposed in the SME.

The Lorentz- and CPT-violating dynamical coefficients are inherently sourced by the presence of non-trivial non-metricity. { Unlike} in \cite{Delhom:2019gxg}, where the authors focused only on the first-order contributions in $\xi$, here we have obtained the spinor action in the Einstein frame to all orders in $\xi$. By expanding this action  in a power series of $\xi$, we have shown that the results match those found in \cite{Delhom:2019gxg} in the first order of approximation, though we also see that axial contributions arise in the second order in $\xi$ even in the absence of torsion. Therefore, this indicates that torsion-like terms might be emulated by purely non-metricity ones. Of course, these ``anomalous'' contributions can be neglected by requiring that $\xi$  be Planck-suppressed (which is reasonable in order to avoid the emergence of undesirable instabilities as exhaustively discussed in \cite{Delhom:2019gxg}), and then only the leading order effectively contributes.

We have considered quantum aspects in this scenario by calculating the one-loop effective action in the Einstein frame. To do that, we have used functional methods, namely, the Barvinsky-Vilkovisky technique, to obtain the divergent part of the one-loop effective action, i.e., the fermionic determinant. In practical terms, this determinant is split into two parts: the first one comes from the minimal contributions to the quadratic action while the second stems from the non-minimal contributions of the quadratic action. By doing so, we were able to find the divergent part of the full one-loop effective quantum gravity action given by Eq.~(\ref{434}). Under these considerations, we have concluded that, clearly, the resulting theory has renormalizability and/or unitarity problems, as expected in the Einstein frame representation. In the weak field approximation (Minkowskian effective theory), such issues can be left aside because the divergences of the one-loop effective action dramatically simplify, as it boils down to just two terms (see Eq.~(\ref{Gd2})). The first term contributes to the vacuum energy, while the second one corresponds to Fig. \ref{fig:fig1}. It is worth noting that this final result matches the calculations performed in \cite{pj20} with the use of Feynman diagrams.

A natural continuation of this work involves  the study of two different aspects of the theory: classical and quantum ones. In the former, we intend to study the impact of the non-linear LV coefficients found here in cosmological and astrophysical backgrounds. In the latter, a detailed study of the spontaneous Lorentz symmetry breaking in an effective theory whose action includes the one-loop contributions computed here. Another interesting problem consists in calculating the finite part of the one-loop correction. These analyses are currently underway. 

\appendix

\appendix
\section{Dirac action in the Einstein frame}
\label{appendix3}

Here we will show the steps that have to be in order to write the fermionic Einstein frame action only in terms of the dynamic fields of the Einstein frame. To that end, we must first find a solution for the connection which, as we will see, is an auxiliary field that can be written algebraically in terms of $h$, $\partial h$ $B$ and $\Psi$ according to its field equations. We will then plug this solution back in the action and find the sought Einstein frame form of the Dirac action \eqref{ferm} in terms of $h$, $B$ and $\Psi$. Through this appendix all tangent space (greek) indices are risen and lowered with $h_{\mu\nu}$ and mapped to anholonomic frames with the vierbeins $E^a{}_\mu$, so we will omit the tildes to lighten the notation. 

To do that, note that a splitting of the connection of the form $\Gamma^\lambda{}_{\mu\nu}={}^h\Gamma^\lambda{}_{\mu\nu}+\Upsilon^\lambda_{\mu\nu}$, where ${}^h\Gamma^\lambda_{\mu\nu}$ is the Levi-Civita connection of $h_{\mu\nu}$, leads to a splitting in the spin connection \eqref{gamu} which, after using the above relations \eqref{vierrelations} for the vierbeins reads
\begin{eqnarray}
\omega^{(\Gamma)}_{\mu ab}&=&\;\omega^{(h)}_{\mu ab}+\frac{1}{2}\lr{\eta_{ab}-\xxsp\theta_{ab}} \partial_\mu\ln \xx+\xxsp\left[\xx^{-1}\eta_{ac}E^c{}_\nu\nablah_\mu\theta^\nu{}_b\right.\nonumber\\ &-&
\left.
\theta_{a\nu}\left(\nablah_\mu E^\nu{}_b+\xxsm\nablah_\mu\theta^\nu{}_b\right)\right]\\
&+&\lrsq{\eta_{ac}E^c{}_\nu E^\beta{}_b+\xxsp\lr{\xx^{-1}\eta_{ac}E^c{}_\nu\theta^\beta{}_b-\theta_{a\nu}\left(E^\beta{}_b+\xxsm\theta^\beta{}_b\right)}}\Upsilon^\nu_{\mu\beta}\equiv\omega^{(h)}_{\mu ab}+\delta\omega_{\mu ab}.\nonumber
\label{gamusplit}
\end{eqnarray}
where $\omega^{(h)}_{\mu ab}=\eta_{ac}E^c{}_\nu\lr{\partial_\mu E^\nu{}_b+E^\beta{}_b{}^h\Gamma^\nu_{\mu\beta}}$ we have defined $\xx=(1+\xi X)^{1/2}$ and $\theta^a_b=B^aB_b$ to lighten the notation. In our case, when the field equations of the connection are given by Eq. \eqref{connectioneq}, $\Upsilon$ is related to the hypermomentum as in Eq. \eqref{Upsilon}. To find the explicit solution to the connection, we must then calculate the explicit form of the hypermomentum and then solve Eq. \eqref{Upsilon} for $\Upsilon^\lambda_{\mu\nu}$. In order to do that, recall that from the definition of the hypermomentum, it concerns variations of the matter Lagrangian with respect to the connection when all other fields are constant, so  that we can write
\begin{equation}
\Delta_{\lambda}^{\mu\nu} \equiv\left. 2 \frac{\delta\left(\sqrt{-g}\mathcal{L}_M \right)}{\delta \Gamma^{\lambda}_{\mu\nu}}\right|_{g,B,\Psi}= 2 \frac{\delta \left(\sqrt{-g}\mathcal{L}_M \right)}{\delta \Upsilon^{\lambda}_{\mu\nu}}.
\end{equation}
By using the definition of the spinor covariant derivative \eqref{spinorcovder}, the splitting of the spin connection into its Levi-Civita part and the rest \eqref{gamusplit}, and the identity involving Dirac matrices $\{\gamma^a,[\gamma^b,\gamma^c]\}=4i\epsilon^{abcd}\gamma_d\gamma_5$, we can rewrite the spinor matter action \eqref{ferm} as 
\begin{equation}
\begin{split}
\mbox{S}_{f}=\int d^{4}x&\,\sqrt{-h}\Bigg\{\frac{i}{2}\xx^{-1/2}E_{\,\,a}^{\mu}\left(\bar{\Psi}\gamma^{a}(\nablah_{\mu}\Psi)-(\nablah_{\mu}\bar{\Psi})\gamma^a\Psi\right)-\xx^{-1}m\bar{\Psi}\Psi\\
&+\xx^{1/2}\theta^\mu_a\left(\bar{\Psi}\gamma^{a}(\nablah_{\mu}\Psi)-(\nablah_{\mu}\bar{\Psi})\gamma^{a}\Psi\right)-\frac{1}{4}\lr{E_{\,\,a}^{\mu}+\xx^{1/2}\theta^\mu_a}\delta\omega_{\mu bc}\epsilon^{abcd}\ja_d,
\label{intermDirac}
\end{split}
\end{equation}
where $\ja^a=\bar{\Psi}\gamma^a\gamma_5\Psi$ is the axial fermionic current. In order to compute the fermionic hypermomentum, note that the only piece of the above action that depends on $\Upsilon^\alpha_{\mu\nu}$ is the last one, so that the hypermomentum in the Einstein frame reads
\begin{equation}
\Delta_\lambda^{\mu\nu}=-\frac{\sqrt{-h}}{2}\lr{E_{\,\,a}^{\alpha}+\xx^{1/2}\theta^\alpha_a}\epsilon^{abcd}\ja_d\frac{\partial \delta\omega_{\alpha bc}}{\partial\Upsilon^\lambda_{\mu\nu}}
\end{equation}
which leads to
\begin{eqnarray}
\Delta_\lambda^{\mu\nu}&=&-\frac{\sqrt{-h}}{2}\lr{E_{\,\,a}^{\mu}+\xx^{1/2}\theta^\mu_a}\epsilon^{abcd}\ja_d
\lrsq{\eta_{bk}E^k{}_\lambda E^\nu{}_c+\xxsm\eta_{bk}E^c{}_\lambda\theta^\nu{}_c-\right.\nonumber\\ &-&\left.
	\theta_{b\lambda}\left(\xxsp E^\nu{}_c+\theta^\nu{}_c\right)}
\end{eqnarray}
where the first term contains the hypermomentum fermions in metric-affine GR proportional to the axial current and a Levi-Civita symbol, and the rest is due to the presence of the nonminimal coupling between the bumblebee and the Ricci tensor. In order to provide an explicit solution for the connection, we must be able to isolate $\Upsilon^\lambda_{\mu\nu}$ from Eq. \eqref{Upsilon}. To do that, note that taking the trace of Eq. \eqref{Upsilon} with $\delta^\beta_\kappa$, and noting that all traces of the above hypermomentum vanish, we find that $\Upsilon^\alpha_{\alpha\gamma}=0$, and using this result, Eq. \eqref{Upsilon} yields
\begin{eqnarray}
		\Upsilon^\kappa_{\beta\gamma}&=&\kappa^2\epsilon^{abcd}\ja_d h^{\kappa\lambda}\Bigg[\lr{h_{\gamma\mu}E_{\,\,a}^{\mu}+\xx^{1/2}\theta_{\gamma a}}\lrsq{\eta_{bk}E^k{}_\beta h_{\lambda\nu} E^\nu{}_c+\xxsm\eta_{bk}E^c{}_\beta\theta_{\lambda c}-\right.\nonumber\\
			&&\left.-
			\theta_{b\beta}\left(\xxsp h_{\lambda\nu}E^\nu{}_c+\theta_{\lambda c}\right)}\nonumber\\
		&&+\lr{h_{\lambda\mu}E_{\,\,a}^{\mu}+\xx^{1/2}\theta_{\lambda a}}\lrsq{\eta_{bk}E^k{}_\gamma h_{\beta\nu} E^\nu{}_c+\xxsm\eta_{bk}E^c{}_\gamma\theta_{\beta c}-\theta_{b\gamma}\left(\xxsp h_{\beta\nu}E^\nu{}_c+\theta_{\beta c}\right)}\nonumber\\
		&&+\lr{h_{\beta\mu}E_{\,\,a}^{\mu}+\xx^{1/2}\theta_{\beta a}}\lrsq{\eta_{bk}E^k{}_\lambda h_{\gamma\nu} E^\nu{}_c+\xxsm\eta_{bk}E^c{}_\lambda\theta_{\gamma c}-\theta_{b\lambda}\left(\xxsp h_{\gamma\nu}E^\nu{}_c+\theta_{\gamma c}\right)}\Bigg] \nonumber\\
		&& 
		\equiv \kappa^2\Omega^\kappa_{\beta\gamma}.\label{explicithypermom}
\end{eqnarray}

 Having solved the connection, we can now finish our task of writing the Einstein frame version of the Dirac action in terms of the Einstein frame propagating fields. We do that by expanding the $\delta\omega_{\mu bc}$ term in Eq. \eqref{intermDirac}, and keeping track of the terms that vanish due to contracting with $\epsilon^{abcd}$ (note the symmetries of $\theta_{ab}$ and $\theta_{ab}\theta_{cd}$), we arrive at 
 
\begin{equation}
\begin{split}
\mbox{S}_{f}=&\int d^{4}x\,\sqrt{-h}\Bigg\{\frac{i}{2}\xx^{-1/2}E_{\,\,a}^{\mu}\left(\bar{\Psi}\gamma^{a}(\nablah_{\mu}\Psi)-(\nablah_{\mu}\bar{\Psi})\gamma^a\Psi\right)-\xx^{-1}m\bar{\Psi}\Psi\\
&+i\xx^{1/2}\theta^\mu_a\left(\bar{\Psi}\gamma^{a}(\nablah_{\mu}\Psi)-(\nablah_{\mu}\bar{\Psi})\gamma^{a}\Psi\right)-\frac{\xxsp}{4}\epsilon^{abcd}\ja_d\Bigg[\xx^{-1}\eta_{ak}E^\mu{}_c E^{k}{}_\nu\nablah_\mu\theta^\nu_b\\
&-\theta_{a\nu}E^\mu{}_c\lr{\nablah_\mu E^\nu{}_b+\xxsm \nablah_\mu\theta^\nu{}_b}\Bigg]-\kappa^2\Sigma(h,B,\ja)\Bigg\}\\
\end{split}
\end{equation}
where $\Sigma$ comes from the hypermomentum contribution to $\delta\omega_{\mu bc}$ and is given by
\begin{eqnarray}
	\Sigma&=&\frac{E_{\,\,a}^{\mu}+\xx^{1/2}\theta^\mu_a}{4}\epsilon^{abcd}\ja_d\Omega^\nu_{\mu\beta} 
	\times\nonumber\\&\times&
	\lrsq{\eta_{bk}E^k{}_\nu E^\beta{}_c+\xxsp\lr{\xx^{-1}\eta_{bk}E^k{}_\nu\theta^\beta{}_c-\theta_{b\nu}\left(E^\beta{}_c+\xxsm\theta^\beta{}_c\right)}}.
\end{eqnarray}
Expanding now the remaining $\theta_{ab}$ terms in the above action, taking into account vanishing terms due to contraction with the Levi-Civita symbol, and integrating out a boundary term, we are led to
\begin{equation}
\begin{split}
\mbox{S}_{f}=&\int d^{4}x\,\sqrt{-h}\Bigg\{iE_{\,\,a}^{\mu}\bar{\Psi}\gamma^{a}\nablah_{\mu}\Psi-m\frac{1}{(1+\xi X)^{1/4}}\bar{\Psi}\Psi+i\frac{1-\sqrt{1+\xi X}}{X(1+\xi X)^{3/4}}\tilde{B}^\mu \tilde{B}_a\bar{\Psi}\gamma^{a}\nablah_{\mu}\Psi\\
&+i\frac{1-\sqrt{1+\xi X}}{2X(1+\xi X)^{3/4}}\lr{\tilde{T}_\alpha \bar{\Psi}\tilde{\gamma}^\alpha\Psi+i\tilde{\ja}^\alpha \tilde{S}_\alpha}-\kappa^2\Sigma(h,B,\ja)\Bigg\}\\
\end{split}
\end{equation}
where we have absorbed a $\xx^{-1/2}$ factor into the volume element and defined
\begin{equation}
\begin{split}
{\tilde T}_{\mu}&={\lr{1+\xi X}^{-1/4}E^a{}_\mu}\lr{\nablah_\nu \tilde{B}^\nu B_a},\\
{\tilde S}_{\mu}&={\frac{1}{2 (1+\xi X)^{1/2}}}\epsilon^{abcd}E_{\mu d} E^{\lambda}{}_{b}\tilde{B}_{c}\left[\lr{1-\sqrt{1+\xi X}}E^{\nu}_{a}\nablah_{\lambda}\tilde{B}_{\nu}{+}B_\nu\nablah_\lambda E^{\nu}{}_a\right].
\label{AS}
\end{split}
\end{equation}

\section{Calculation of the coefficient $a_{2}(x,x^{\prime})$}
\label{appendix2}
 This appendix is devoted to demonstrating some steps of the calculation of the divergent part of the one-loop effective action. Let us start substituting Eq. (\ref{def2}) into Eqs. (\ref{p1}, \ref{w1}) to get the parameters:
\begin{equation}
\begin{split}
\hat{P}&=\left(m_{0}^{2}-\frac{\tilde{R}}{12}\right)\hat{1}+\frac{1}{8}\xi \tilde{\gamma}^{\mu}\tilde{\gamma}^{\nu}\tilde{F}_{\mu\nu}+\frac{i}{4}\xi\gamma_{5}\tilde{\nabla}^{\mu}\tilde{S}_{\mu}-\frac{1}{4}\xi \tilde{\gamma}^{\nu}\tilde{\gamma}^{\beta}\tilde{\nabla}_{\mu}\tilde{\nabla}_{\nu}\left(B_{\beta}B^{\mu}\right)+\\
&+\xi B^{\nu}B_{\alpha}\tilde{R}_{\mu\nu}\tilde{\gamma}^{\alpha}\tilde{\gamma}^{\mu}
,\\
\hat{W}_{\mu\nu}&=\frac{1}{4}\tilde{R}_{\mu\nu\alpha\beta}\tilde{\gamma}^{\alpha}\tilde{\gamma}^{\beta}-\frac{1}{4}\xi \tilde{F}_{\mu\nu}+\frac{1}{4}\xi \sigma_{\mu\lambda}\gamma_{5}\tilde{\nabla}_{\nu}\tilde{S}^{\lambda}-\frac{1}{4}\sigma_{\nu\lambda}\gamma_{5}\tilde{\nabla}_{\mu}\tilde{S}^{\lambda}+\\
&+\frac{\xi}{4}\tilde{\gamma}^{\alpha}\tilde{\gamma}^{\beta}\tilde{\nabla}_{\mu}\tilde{\nabla}_{\alpha}(B_{\beta}B_{\nu})-\frac{\xi}{4}\tilde{\gamma}^{\alpha}\tilde{\gamma}^{\beta}\tilde{\nabla}_{\nu}\tilde{\nabla}_{\alpha}(B_{\beta}B_{\mu}).
\end{split}
\end{equation}
Now, contracting these objects, we find:
\begin{equation}
\begin{split}
\hat{P}^{2}&=\left(m_{0}^{4}-\frac{1}{6}m_{0}^{2}\tilde{R}+\frac{1}{144}\tilde{R}^{2}\right)\hat{1}+\frac{1}{4}\xi m_{0}^{2}\tilde{\gamma}^{\mu}\tilde{\gamma}^{\nu}\tilde{F}_{\mu\nu}+\frac{i}{2}\xi m_{0}^{2}\gamma_{5}\tilde{\nabla}^{\mu}\tilde{S}_{\mu}-\frac{1}{48}\xi\tilde{R}\tilde{\gamma}^{\mu}\tilde{\gamma}^{\nu}\tilde{F}_{\mu\nu}-\\
&-\frac{1}{2}\xi m_{0}^2\tilde{\gamma}^{\nu}\tilde{\gamma}^{\beta}\tilde{\nabla}_{\mu}\tilde{\nabla}_{\nu}\left(B_{\beta}B^{\mu}\right)+\frac{1}{24}\xi \tilde{R}\tilde{\gamma}^{\nu}\tilde{\gamma}^{\beta}\tilde{\nabla}_{\mu}\tilde{\nabla}_{\nu}\left(B_{\beta}B^{\mu}\right)-\frac{1}{24}\xi \tilde{R}\gamma_{5}\tilde{\nabla}^{\mu}\tilde{S}_{\mu}+\\
&+2m_{0}^{2}\xi B^{\nu}B_{\alpha}\tilde{R}_{\mu\nu}\tilde{\gamma}^{\alpha}\tilde{\gamma}^{\mu}-\frac{\tilde{R}}{6}\xi B^{\nu}B_{\alpha}\tilde{R}_{\mu\nu}\tilde{\gamma}^{\alpha}\tilde{\gamma}^{\mu}
,\\
\hat{\tilde{W}}^{\mu\nu}\hat{W}_{\mu\nu}&=\frac{1}{16}\tilde{R}_{\mu\nu\alpha\beta}\tilde{R}^{\mu\nu}_{\quad \lambda\theta}\tilde{\gamma}^{\alpha}\tilde{\gamma}^{\beta}\tilde{\gamma}^{\lambda}\tilde{\gamma}^{\theta}+\frac{1}{4}\xi\tilde{R}_{\mu\nu\alpha\beta}\tilde{\sigma}^{\mu\lambda}\tilde{\gamma}^{\alpha}\tilde{\gamma}^{\beta}\gamma_{5}\left(\tilde{\nabla}^{\nu}\tilde{S}_{\lambda}\right)-\frac{1}{8}\xi\tilde{R}_{\mu\nu\alpha\beta}\tilde{F}^{\mu\nu}\tilde{\gamma}^{\alpha}\tilde{\gamma}^{\beta}+\\
&+\frac{1}{8}\xi\tilde{R}_{\mu\nu\alpha\beta}\tilde{\gamma}^{\alpha}\tilde{\gamma}^{\beta}\tilde{\gamma}^{\lambda}\tilde{\gamma}^{\theta}\tilde{\nabla}^{\mu}\tilde{\nabla}_{\lambda}(B_{\theta}B^{\nu})
\end{split}
\end{equation}
From Eq. (\ref{hhhh}) and using the Dirac trace properties, we have:
\begin{equation}
\begin{split}
\mbox{Tr}\left(\hat{P}\right)&=4m_{0}^{2}-\frac{1}{3}\tilde{R}-\frac{1}{2}\xi \tilde{\nabla}_{\mu}\tilde{\nabla}_{\nu}\left(B^{\mu}B^{\nu}\right)+4\xi B^{\nu}B_{\alpha}\tilde{R}^{\alpha}_{\,\,\nu},\\
\frac{1}{2}\mbox{Tr}\left(\hat{P}^{2}\right)&=2m_{0}^4-\frac{1}{3}m_{0}^{2}\tilde{R}+\frac{1}{72}\tilde{R}^{2}-\xi m_{0}^2 \tilde{\nabla}_{\mu}\tilde{\nabla}_{\nu}\left(B^{\mu}B^{\nu}\right)+\frac{1}{12}\xi\tilde{R}\tilde{\nabla}_{\mu}\tilde{\nabla}_{\nu}\left(B^{\mu}B^{\nu}\right)+\\
&+4m_{0}^{2}\xi B^{\nu}B_{\alpha}\tilde{R}^{\alpha}_{\,\,\nu}-\frac{\tilde{R}}{3}\xi B^{\nu}B_{\alpha}\tilde{R}^{\alpha}_{\,\,\nu},\\
\frac{1}{12}\mbox{Tr}\left(\hat{\tilde{W}}^{\mu\nu}\hat{W}_{\mu\nu}\right)&=-\frac{1}{24}\tilde{R}_{\mu\nu\lambda\beta}\tilde{R}^{\mu\nu\lambda\beta}+\frac{1}{12}\xi\tilde{R}_{\mu\nu\theta\lambda}\tilde{\nabla}^{\mu}\tilde{\nabla}^{\lambda}\left(B^{\theta}B^{\nu}\right)\\
\end{split}
\end{equation}
This expression is used in our studies in the section 4.

\section{Computation of the nonminimal contributions to the divergent piece of the one-loop effective action}
\label{appendix33}

In this Appendix we provide the explicit calculation of the second piece of Eq. (\ref{effac}) which involves the non-minimal operator. The first step in order to calculate it is to find the inverse operator $\hat{H}^{-1}_{0}$ which has been calculated in \cite{Netto:2014faa} for the gauge sector. Here, we will just adapt it to our case. Thus, the inverse operator can be expanded in inverse power series of $\tilde{\square}$, i.e.,
\begin{equation}
\begin{split}
\hat{H}^{-1}_{0}&=\frac{1}{\tilde{\square}}+\frac{\tilde{R}}{4}\frac{1}{\tilde{\square}^{2}}-\frac{1}{2}\left(\tilde{\nabla}^{\rho}\tilde{R}\right)\tilde{\nabla}_{\rho}\frac{1}{\tilde{\square}^{3}}-\frac{1}{4}\left(\tilde{\square}\tilde{R}\right)\frac{1}{\tilde{\square}^{3}}+\\
&+\frac{\tilde{R}^{2}}{4}\frac{1}{\tilde{\square}^{3}}+\left(\tilde{\nabla}^{\rho}\tilde{\nabla}^{\sigma}\tilde{R}\right)\tilde{\nabla}_{\rho}\tilde{\nabla}_{\sigma}\frac{1}{\tilde{\square}^{4}}+...
\end{split}
\end{equation}  
where the ellipsis stand for irrelevant terms.

Now, one can compute the divergences of the second piece of one-loop effective action that takes the form
\begin{equation}
\begin{split}
-\frac{i}{2}\mbox{Tr}\left(\hat{H}_{nm}\hat{H}^{-1}_{0}\right)&=-\frac{i}{2}\xi\,\mbox{Tr}\bigg\{B^{\mu}B^{\nu}\bigg[\tilde{\nabla}_{\mu}\tilde{\nabla}_{\nu}\frac{1}{\tilde{\square}}+\frac{1}{4}\left(\tilde{\nabla}_{\mu}\tilde{\nabla}_{\nu}\tilde{R}\right)\frac{1}{\tilde{\square}^{2}}+\\
&+\frac{1}{4}\tilde{R}\tilde{\nabla}_{\mu}\tilde{\nabla}_{\nu}\frac{1}{\tilde{\square}^{2}}+\frac{1}{2}\left(\tilde{\nabla}_{\mu}\tilde{R}\right)\tilde{\nabla}_{\nu}\frac{1}{\tilde{\square}^{2}}+\\
&+\frac{1}{4}\tilde{R}^{2}\tilde{\nabla}_{\mu}\tilde{\nabla}_{\nu}\frac{1}{\tilde{\square}^{3}}-\frac{1}{2}\left(\tilde{\nabla}^{\rho}\tilde{R}\right)\tilde{\nabla}_{\mu}\tilde{\nabla}_{\nu}\tilde{\nabla}_{\rho}\frac{1}{\tilde{\square}^{3}}-\\
&-\left(\tilde{\nabla}_{\mu}\tilde{\nabla}^{\rho}\tilde{R}\right)\tilde{\nabla}_{\nu}\tilde{\nabla}_{\rho}\frac{1}{\tilde{\square}^{3}}-\frac{1}{4}\left(\tilde{\square}\tilde{R}\right)\tilde{\nabla}_{\mu}\tilde{\nabla}_{\nu}\frac{1}{\tilde{\square}^{3}}+\\
&+\left(\tilde{\nabla}^{\rho}\tilde{\nabla}^{\sigma}\tilde{R}\right)\tilde{\nabla}_{\mu}\tilde{\nabla}_{\nu}\tilde{\nabla}_{\rho}\tilde{\nabla}_{\sigma}\frac{1}{\tilde{\square}^{4}}\bigg]\bigg\}.
\label{nm}
\end{split}
\end{equation} 

Each term in the former equation can be computed by using the table of the universal traces displayed in \cite{Barvinsky:1985an}. Following this table, we find 
\begin{equation}
\begin{split}
-i\frac{\xi}{8}\mbox{Tr}B^{\mu}B^{\nu}\left(\tilde{\nabla}_{\mu}\tilde{\nabla}_{\nu}\tilde{R}\right)\frac{1}{\tilde{\square}^{2}}\Big|_{\mbox{div}}&=-\frac{\xi\,\mu^{D-4}}{(4\pi)^{2}\epsilon}\int d^{D}x\sqrt{-h}\,B^{\mu}B^{\nu}\tilde{\nabla}_{\mu}\tilde{\nabla}_{\nu}\tilde{R},
\label{first}
\end{split}
\end{equation}

\begin{equation}
\begin{split}
-i\frac{\xi}{2}\mbox{Tr}B^{\mu}B^{\nu}\tilde{\nabla}_{\mu}\tilde{\nabla}_{\nu}\frac{1}{\tilde{\square}}\Big|_{\mbox{div}}&=\frac{\xi\,\mu^{D-4}}{4(4\pi)^{2}\epsilon}\int d^{D}x\sqrt{-h}\,\bigg\{\frac{4}{45}B^{\mu}B^{\nu}\tilde{R}^{\alpha\beta}\tilde{R}_{\alpha\mu\beta\nu}-\frac{8}{45}B^{\mu}B^{\nu}\tilde{R}_{\mu\alpha}\tilde{R}^{\alpha}_{\,\,\nu}+\\
&+\frac{2}{9}B^{\mu}B^{\nu}\tilde{R}\tilde{R}_{\mu\nu}+\frac{2}{15}B^{\mu}B^{\nu}\tilde{\square}\tilde{R}_{\mu\nu}+\frac{2}{5}B^{\mu}B^{\nu}\tilde{\nabla}_{\mu}\tilde{\nabla}_{\nu}\tilde{R}+\\
&+X\left(\frac{7}{360}\tilde{R}_{\mu\nu\alpha\beta}\tilde{R}^{\mu\nu\alpha\beta}+\frac{1}{45}\tilde{R}_{\mu\nu}R^{\mu\nu}-\frac{1}{72}\tilde{R}^2-\frac{2}{15}\tilde{\square}\tilde{R}\right)-\\
&-\frac{7}{90}B^{\alpha}B^{\beta}\tilde{R}_{\alpha\lambda\mu\nu}\tilde{R}^{\,\,\,\lambda\mu\nu}_{\beta}
\bigg\},
\end{split}
\end{equation}

\begin{equation}
\begin{split}
-i\frac{\xi}{8}\mbox{Tr}\tilde{R}B^{\mu}B^{\nu}\tilde{\nabla}_{\mu}\tilde{\nabla}_{\nu}\frac{1}{\tilde{\square}^{2}}\Big|_{\mbox{div}}&=-\frac{\xi\,\mu^{D-4}}{(4\pi)^{2}\epsilon}\int d^{D}x\sqrt{-h}\bigg[\frac{1}{6}\tilde{R}B^{\mu}B^{\nu}\tilde{R}_{\mu\nu}-\frac{1}{12}X\tilde{R}^2\bigg],
\end{split}
\end{equation}

\begin{equation}
\begin{split}
-i\frac{\xi}{8}\mbox{Tr}B^{\mu}B^{\nu}\tilde{R}^{2}\tilde{\nabla}_{\mu}\tilde{\nabla}_{\nu}\frac{1}{\tilde{\square}^{3}}\Big|_{\mbox{div}}&=-\frac{\xi\,\mu^{D-4}}{4(4\pi)^{2}\epsilon}\int d^{D}x\sqrt{-h}\,X\tilde{R}^2,
\end{split}
\end{equation}

\begin{equation}
\begin{split}
i\frac{\xi}{8}\mbox{Tr}B^{\mu}B^{\nu}\left(\tilde{\square} \tilde{R}\right)\tilde{\nabla}_{\mu}\tilde{\nabla}_{\nu}\frac{1}{\tilde{\square}^{3}}\Big|_{\mbox{div}}&=\frac{\xi\,\mu^{D-4}}{4(4\pi)^{2}\epsilon}\int d^{D}x\sqrt{-h}\,X\tilde{\square}\tilde{R},
\end{split}
\end{equation}

\begin{equation}
\begin{split}
i\frac{\xi}{2}\,\mbox{Tr}B^{\mu}B^{\nu}\left(\tilde{\nabla}_{\mu}\tilde{\nabla}^{\rho}\tilde{R}\right)\tilde{\nabla}_{\nu}\tilde{\nabla}_{\rho}\frac{1}{\tilde{\square}^{3}}\Big|_{\mbox{div}}&=\frac{\xi\,\mu^{D-4}}{(4\pi)^{2}\epsilon}\int d^{D}x\sqrt{-h}\,B^{\mu}B_{\rho}\tilde{\nabla}_{\mu}\tilde{\nabla}^{\rho}\tilde{R},
\end{split}
\end{equation}

\begin{equation}
\begin{split}
&-i\frac{\xi}{2}\,\mbox{Tr}B^{\mu}B^{\nu}\left(\tilde{\nabla}^{\rho}\tilde{\nabla}^{\sigma}\tilde{R}\right)\tilde{\nabla}_{\mu}\tilde{\nabla}_{\nu}\tilde{\nabla}_{\rho}\tilde{\nabla}_{\sigma}\frac{1}{\tilde{\square}^{4}}\Big|_{\mbox{div}}=\\
&-\frac{\xi\,\mu^{D-4}}{3(4\pi)^{2}\epsilon}\int d^{D}x\sqrt{-h}\,B_{\mu}B_{\nu}\tilde{\nabla}^{\mu}\tilde{\nabla}^{\nu}\tilde{R}-\frac{\xi\,\mu^{D-4}}{6(4\pi)^{2}\epsilon}\int d^{D}x\sqrt{-h}\,X\tilde{\square}\tilde{R},
\end{split}
\end{equation}

\begin{equation}
\begin{split}
2\left(\tilde{\nabla}_{\mu}\tilde{R}\right)\tilde{\nabla}_{\nu}\frac{1}{\tilde{\square}^{2}}\Big|_{\mbox{div}}=-2\left(\tilde{\nabla}^{\rho}\tilde{R}\right)\tilde{\nabla}_{\mu}\tilde{\nabla}_{\nu}\tilde{\nabla}_{\rho}\frac{1}{\tilde{\square}^{3}}\Big|_{\mbox{div}}=0.
\label{last}
\end{split}
\end{equation}

\acknowledgments
This work was partially supported by Conselho Nacional de Desenvolvimento Cient\'{\i}fico e Tecnol\'{o}gico (CNPq) and by the Spanish Grants FIS2017-84440-C2-1-P and PID2020-116567GB-C21 funded by MCIN/AEI/10.13039/501100011033 (“ERDF A way of making Europe”), and the project PROMETEO/2020/079 (Generalitat Valenciana). The work by A. Yu. P. has been supported by the CNPq project No. 301562/2019-9. PJP would like to thank the Brazilian agency CAPES for financial support (PNPD/CAPES grant, process 88887.464556/2019-00) and Department de Física Teòrica and IFIC, Universitat de València, for hospitality. AD gratefully acknowledges the full support by the Estonian Research Council and the European Regional Development Fund through the grant Center of Excellence TK133 “The Dark Side of the Universe”. AD also wants to thank for hospitality to the Departamento de Física da Universidade Federal da Paraíba.




\begin{thebibliography}{99}


\bibitem{datatables} V.~A.~Kostelecky and N.~Russell,
\textit{Data Tables for Lorentz and CPT Violation},
arXiv:0801.0287 [hep-ph].

\bibitem{Kostelecky:1988zi}                    
V.~A.~Kostelecky and S.~Samuel,
\textit{Spontaneous Breaking of Lorentz Symmetry in String Theory},
Phys. Rev. D \textbf{39}, 683 (1989).

\bibitem{Kostelecky:1989jp}
V.~A.~Kostelecky and S.~Samuel,
\textit{Phenomenological Gravitational Constraints on Strings and Higher Dimensional Theories},
Phys. Rev. Lett. \textbf{63}, 224 (1989).

\bibitem{Kostelecky:1989jw}
V.~A.~Kostelecky and S.~Samuel,
\textit{Gravitational Phenomenology in Higher Dimensional Theories and Strings},     
Phys. Rev. D \textbf{40}, 1886-1903 (1989).

\bibitem{Kostelecky:1991ak}
V.~A.~Kostelecky and R.~Potting,
\textit{CPT and strings},
Nucl. Phys. B \textbf{359}, 545-570 (1991).

\bibitem{Kostelecky:1994rn}
V.~A.~Kostelecky and R.~Potting,
\textit{CPT, strings, and meson factories},
Phys. Rev. D \textbf{51}, 3923-3935 (1995).

\bibitem{Georgi} H. Georgi, \textit{Effective field theory}, Ann. Rev. Nucl. Part. Sci. {\bf 43}, 209 (1993).

\bibitem{Colladay:1996iz}                    
D.~Colladay and V.~A.~Kostelecky,
\textit{CPT violation and the standard model},
Phys. Rev. D \textbf{55}, 6760-6774 (1997)          
[arXiv:hep-ph/9703464 [hep-ph]].

\bibitem{Colladay:1998fq}
D.~Colladay and V.~A.~Kostelecky,
\textit{Lorentz violating extension of the standard model},   
Phys. Rev. D \textbf{58}, 116002 (1998)
[arXiv:hep-ph/9809521 [hep-ph]].

\bibitem{Kostelecky:2003fs}
V.~Kostelecky, \textit{Gravity, Lorentz violation, and the standard model},
Phys. Rev. D \textbf{69} (2004), 105009,
[arXiv:hep-th/0312310 [hep-th]].


\bibitem{Jacobson:2000xp}
T.~Jacobson and D.~Mattingly,
\textit{Gravity with a dynamical preferred frame},
Phys. Rev. D \textbf{64}, 024028 (2001)
[arXiv:gr-qc/0007031 [gr-qc]].

\bibitem{KosLi} V.~A.~Kostelecky and Z.~Li,
Phys.\ Rev.\ D {\bf 103}, 024059 (2021)
[arXiv:2008.12206 [gr-qc]].

\bibitem{Carroll:1989vb} 
S.~M.~Carroll, G.~B.~Field and R.~Jackiw,
\textit{Limits on a Lorentz and Parity Violating Modification of Electrodynamics},
Phys. Rev. D \textbf{41} (1990), 1231.

\bibitem{Jackiw:2003pm}
R.~Jackiw and S.~Y.~Pi,
\textit{Chern-Simons modification of general relativity},
Phys. Rev. D \textbf{68}, 104012 (2003)
[arXiv:gr-qc/0308071 [gr-qc]].

\bibitem{Mirzagholi:2020irt}
L.~Mirzagholi, E.~Komatsu, K.~D.~Lozanov and Y.~Watanabe,
\textit{Effects of Gravitational Chern-Simons during Axion-SU(2) Inflation},
JCAP \textbf{06}, 024 (2020)
[arXiv:2003.05931 [gr-qc]].

\bibitem{Maluf:2021lwh}                
R.~V.~Maluf and J.~C.~S.~Neves,
\textit{Bumblebee field as a source of cosmological anisotropies},
JCAP \textbf{10}, 038 (2021)
[arXiv:2105.08659 [gr-qc]].

\bibitem{KumarJha:2020ivj}
S.~Kumar Jha, H.~Barman and A.~Rahaman,
\textit{Bumblebee gravity and particle motion in Snyder noncommutative spacetime structures},
JCAP \textbf{04}, 036 (2021)
[arXiv:2012.02642 [hep-th]].

\bibitem{Casana:2017jkc}
R.~Casana, A.~Cavalcante, F.~P.~Poulis and E.~B.~Santos,
\textit{Exact Schwarzschild-like solution in a bumblebee gravity model},
Phys. Rev. D \textbf{97}, no.10, 104001 (2018)
[arXiv:1711.02273 [gr-qc]].

\bibitem{Gullu:2020qzu}
\.I.~G\"ull\"u and A.~\"Ovg\"un,
\textit{Schwarzschild-like black hole with a topological defect in bumblebee gravity},
Annals Phys. \textbf{436}, 168721 (2022)
[arXiv:2012.02611 [gr-qc]].

\bibitem{Maluf:2020kgf}                          
R.~V.~Maluf and J.~C.~S.~Neves,
\textit{Black holes with a cosmological constant in bumblebee gravity},
Phys. Rev. D \textbf{103}, no.4, 044002 (2021)
[arXiv:2011.12841 [gr-qc]].

\bibitem{Bartolo:2017szm}
N.~Bartolo and G.~Orlando,
\textit{Parity breaking signatures from a Chern-Simons coupling during inflation: the case of non-Gaussian gravitational waves},
JCAP \textbf{07}, 034 (2017)
[arXiv:1706.04627 [astro-ph.CO]].

\bibitem{Conroy:2019ibo}
A.~Conroy and T.~Koivisto,
\textit{Parity-Violating Gravity and GW170817 in Non-Riemannian Cosmology},
JCAP \textbf{12}, 016 (2019)
[arXiv:1908.04313 [gr-qc]].

\bibitem{Boudet:2022wmb}
S.~Boudet, F.~Bombacigno, G.~J.~Olmo and P.~J.~Porfirio,
{\it Quasinormal modes of Schwarzschild black holes in projective invariant Chern-Simons modified gravity}, 
JCAP \textbf{05}, 032 (2022) 
[arXiv:2203.04000 [gr-qc]].



\bibitem{Gao:2020qxy}
X.~Gao,
\textit{Higher derivative scalar-tensor theory from the spatially covariant gravity: a linear algebraic analysis},
JCAP \textbf{11}, 004 (2020)
[arXiv:2006.15633 [gr-qc]].



\bibitem{ourtorsion}
J.~R.~Nascimento, A.~Y.~Petrov and P.~J.~Porf\'\i{}rio,
\textit{Induced gravitational topological term and the Einstein-Cartan modified theory},
Phys. Rev. D \textbf{105}, no.4, 044053 (2022)
[arXiv:2108.05705 [gr-qc]].


\bibitem{Li:2020xjt}
M.~Li, H.~Rao and D.~Zhao,
\textit{A simple parity violating gravity model without ghost instability},
JCAP \textbf{11}, 023 (2020)
[arXiv:2007.08038 [gr-qc]].

\bibitem{Olmo:2011uz}
G.~J.~Olmo,
\textit{Palatini Approach to Modified Gravity: f(R) Theories and Beyond},
Int. J. Mod. Phys. D \textbf{20} (2011), 413-462,
[arXiv:1101.3864 [gr-qc]].

\bibitem{Delhom:2019gxg}    
A.~Delhom, J.~Nascimento, G.~J.~Olmo, A.~Y.~Petrov and P.~Porf\'{i}rio, \textit{Metric-affine bumblebee gravity: classical aspects},
Eur.Phys.J. C \textbf{81} (2021), 287
[arXiv:1911.11605 [hep-th]].

\bibitem{pj20}
A.~Delhom, J.~Nascimento, G.~J.~Olmo, A.~Y.~Petrov and P.~Porf\'{i}rio, \textit{Radiative corrections in metric-affine bumblebee model},
Phys. Lett. B \textbf{826} (2022) 136932
[arXiv: 2010.06391 [hep-th]].




\bibitem{Afonso:2017bxr}
V.~I.~Afonso, C.~Bejarano, J.~Beltran Jimenez, G.~J.~Olmo and E.~Orazi,
\textit{The trivial role of torsion in projective invariant theories of gravity with non-minimally coupled matter fields},
Class. Quant. Grav. \textbf{34} (2017), 235003
[arXiv:1705.03806 [gr-qc]].

\bibitem{BeltranJimenez:2017doy}
J.~Beltran Jimenez, L.~Heisenberg, G.~J.~Olmo and D.~Rubiera-Garcia,
\textit{Born\textendash{}Infeld inspired modifications of gravity,}
Phys. Rept. \textbf{727} (2018), 1-129
[arXiv:1704.03351 [gr-qc]].

\bibitem{Delhom:2022vba}
A.~Delhom,
\textit{Theoretical and Observational Aspecs in Metric-Affine Gravity: A field theoretic perspective},
[arXiv:2201.09789 [gr-qc]].

\bibitem{BeltranJimenez:2019acz}
J.~Beltr\'an Jim\'enez and A.~Delhom,
\textit{Ghosts in metric-affine higher order curvature gravity,}
Eur. Phys. J. C \textbf{79}, no.8, 656 (2019)
[arXiv:1901.08988 [gr-qc]].

\bibitem{BeltranJimenez:2020sqf}
J.~Beltr\'an Jim\'enez and A.~Delhom,
\textit{Instabilities in metric-affine theories of gravity with higher order curvature terms,}
Eur. Phys. J. C \textbf{80} (2020) no.6, 585
[arXiv:2004.11357 [gr-qc]].

\bibitem{BeltranJimenez:2019hrm}
J.~Beltr\'an Jim\'enez and F.~J.~Maldonado Torralba,
\textit{Revisiting the stability of quadratic Poincar\'e gauge gravity,}
Eur. Phys. J. C \textbf{80}, no.7, 611 (2020)
[arXiv:1910.07506 [gr-qc]].

\bibitem{Percacci:2020ddy}
R.~Percacci and E.~Sezgin,
\textit{New class of ghost- and tachyon-free metric affine gravities,}
Phys. Rev. D \textbf{101}, no.8, 084040 (2020)
[arXiv:1912.01023 [hep-th]].

\bibitem{Jimenez-Cano:2022sds}
A.~Jim\'enez-Cano and F.~J.~Maldonado Torralba,
\textit{Vector stability in quadratic metric-affine theories,}
[arXiv:2205.05674 [gr-qc]].

\bibitem{Bek} J. D. Bekenstein, \textit{Relativistic gravitation theory for the MOND paradigm}, 
	Phys. Rev. D70
(2004) 083509, astro-ph/0403694.

\bibitem{Seifert:2009gi}
M.~D.~Seifert,
\textit{Generalized bumblebee models and Lorentz-violating electrodynamics},    
Phys. Rev. D \textbf{81}, 065010 (2010)
[arXiv:0909.3118 [hep-ph]].

\bibitem{Ramazanoglu:2017xbl}
F.~M.~Ramazano\u{g}lu, \textit{Spontaneous growth of vector fields in gravity},
Phys. Rev. D \textbf{96} (2017), 064009
[arXiv:1706.01056 [gr-qc]].

\bibitem{Ramazanoglu:2019jrr} 
F.~M.~Ramazano\u{g}lu and K.~İ.~\"Unl\"ut\"urk,
\textit{Generalized disformal coupling leads to spontaneous tensorization},
Phys. Rev. D \textbf{100} (2019), 084026
[arXiv:1910.02801 [gr-qc]].

\bibitem{Cardoso:2020cwo}
V.~Cardoso, A.~Foschi and M.~Zilhao,
\textit{Collective scalarization or tachyonization: when averaging fails},
Phys. Rev. Lett. \textbf{124} (2020), 221104
[arXiv:2005.12284 [gr-qc]].

\bibitem{Delhom:2020hkb}
A.~Delhom,
\textit{Minimal coupling in presence of non-metricity and torsion},
Eur. Phys. J. C \textbf{80} (2020), 728
[arXiv:2002.02404 [gr-qc]].

\bibitem{Kibble:1961ba}
T.~W.~B.~Kibble,
\textit{Lorentz invariance and the gravitational field},
J. Math. Phys. \textbf{2} (1961), 212-221

\bibitem{CarTam} S.~M.~Carroll and H.~Tam,
Phys.\ Rev.\ D {\bf 78}, 044047 (2008)
[arXiv:0802.0521 [hep-ph]]; 
M.~Gomes, J.~R.~Nascimento, A.~Y.~Petrov and A.~J.~da Silva,
Phys.\ Rev.\ D {\bf 81}, 045018 (2010)
[arXiv:0911.3548 [hep-th]].



\bibitem{Bluhm:2007bd}
R.~Bluhm, S.~H.~Fung and V.~A.~Kostelecky,
\textit{Spontaneous Lorentz and Diffeomorphism Violation, Massive Modes, and Gravity},
Phys. Rev. D \textbf{77} (2008), 065020
[arXiv:0712.4119 [hep-th]].

\bibitem{Bluhm:2004ep}    
R.~Bluhm and V.~A.~Kostelecky,
\textit{Spontaneous Lorentz violation, Nambu-Goldstone modes, and gravity},
Phys. Rev. D \textbf{71} (2005), 065008
[arXiv:hep-th/0412320 [hep-th]].

\bibitem{Seifert:2009vr}  
M.~D.~Seifert, \textit{Vector models of gravitational Lorentz symmetry breaking},
Phys. Rev. D \textbf{79} (2009), 124012
[arXiv:0903.2279 [gr-qc]].

\bibitem{Bailey:2006fd}
Q.~G.~Bailey and V.~A.~Kostelecky,
\textit{Signals for Lorentz violation in post-Newtonian gravity},
Phys. Rev. D \textbf{74} (2006), 045001
[arXiv:gr-qc/0603030 [gr-qc]].

\bibitem{HDCS} T.~Mariz, J.~R.~Nascimento and A.~Y.~Petrov,
Phys. Rev. D \textbf{85} (2012), 125003
[arXiv:1111.0198 [hep-th]].

\bibitem{Foster:2016uui}
J.~Foster, V.~A.~Kostelecky and R.~Xu,
\textit{Constraints on Nonmetricity from Bounds on Lorentz Violation},
Phys. Rev. D \textbf{95} (2017), 084033,
[arXiv:1612.08744 [gr-qc]].

\bibitem{ourrev} A.~F.~Ferrari, J.~R.~Nascimento and A.~Y.~Petrov,
Eur.\ Phys.\ J.\ C {\bf 80}, 459 (2020)
[arXiv:1812.01702 [hep-th]].

\bibitem{Barvinsky:1985an}
A.~O.~Barvinsky and G.~A.~Vilkovisky, \textit{The Generalized Schwinger-Dewitt Technique in Gauge Theories and Quantum Gravity},
Phys. Rept. \textbf{119} (1985), 1-74.

\bibitem{DeBerredoPeixoto:2001qm}
G.~De Berredo-Peixoto, \textit{A Note on the heat kernel method applied to fermions},
Mod. Phys. Lett. A \textbf{16} (2001), 2463-2468.

\bibitem{Netto:2014faa}             
T.~de Paula Netto and I.~L.~Shapiro,
\textit{Vacuum contribution of photons in the theory with Lorentz and CPT-violating terms},
Phys. Rev. D \textbf{89} (2014), 104037
[arXiv:1403.3152 [hep-th]].

\bibitem{Buchbinder:2017zaa}
I.~L.~Buchbinder, T.~de Paula Netto and I.~L.~Shapiro,
\textit{Massive vector field on curved background: Nonminimal coupling, quantization, and divergences},
Phys. Rev. D \textbf{95} (2017), 085009
[arXiv:1703.00526 [hep-th]].

\bibitem{Kostelecky:2000mm}
V.~A.~Kostelecky and R.~Lehnert,
\textit{Stability, causality, and Lorentz and CPT violation},
Phys. Rev. D \textbf{63} (2001), 065008,
[arXiv:hep-th/0012060 [hep-th]].

\bibitem{Buchbinder:1992rb}
I.~L.~Buchbinder, S.~D.~Odintsov and I.~L.~Shapiro,
\textit{Effective action in quantum gravity}, CRC Press, 1992.

\bibitem{heatk}
I.~G.~Avramidi, \textit{Heat kernel and quantum gravity}, Lect. Notes Phys. Monogr. \textbf{64} (2000), 1-149.

\bibitem{heatk1}
I.~G.~Avramidi, \textit{Heat kernel method and its applications}, Springer International Publishing, 2015.

\bibitem{Shapiro:2001rz}
I.~L.~Shapiro,
\textit{Physical aspects of the space-time torsion},
Phys. Rept. \textbf{357} (2002), 113, [arXiv:hep-th/0103093 [hep-th]].

\end{thebibliography}
\end{document}